\documentclass[manuscript]{aastex}
\usepackage{natbib}

\def \upd{\mathrm{d}}
\def \2{$_2$}

\begin{document}

\title{Dynamics of atmospheres with a nondilute condensible component}

\author{
Raymond T. Pierrehumbert}

\affil{Department of Physics, University of Oxford, Oxford OX1 3PU, UK }
\email{Raymond.Pierrehumbert@physics.ox.ac.uk}

\and 
\author{ Feng Ding}
\affil{Department of the Geophysical Sciences, University of Chicago, Chicago IL 60637, USA}




\begin{abstract}
The diversity of characteristics for the host of  recently discovered exoplanets opens 
up a great deal of fertile new territory for geophysical fluid dynamics, particularly when the fluid flow
is coupled to novel thermodynamics, radiative transfer or chemistry. In this paper we survey one of these
new areas -- the climate dynamics of atmospheres with a nondilute condensible component, defined as
the situation in which a condensible component of the atmosphere  
makes up a substantial fraction of the atmospheric mass within some layer. Nondilute dynamics can occur for a wide
range of condensibles, generically applying near both the inner and outer edge of the conventional habitable
zone and in connection with runaway greenhouse phenomena,  but also applies in a wide variety of other planetary circumstances. 
We first present a number of analytical results developing some key features of nondilute atmospheres, and then show
how some of these features are manifest in simulations with a general circulation model adapted to handle
nondilute atmospheres.  We find that nondilute atmospheres have weak horizontal temperature
gradients even for rapidly rotating planets, and that their circulations are largely barotropic. 
The relative humidity of the condensible
component tends towards 100\% as the atmosphere becomes more nondilute, which has important implications for runaway
greenhouse thresholds.  Nondilute atmospheres exhibit a number of interesting organized convection features, for which
there is not yet any adequate theoretical understanding. 
\end{abstract}
%


\section{Introduction}
\label{section:Intro}

One of the most remarkable scientific developments of the past two decades is the
emergence of a capability to detect and characterize planets outside our own 
Solar System.  This has lead to a growing catalogue of exoplanets, currently in
the thousands \citep{batalha2014exploring}. This catalogue is abundant in planets
of a type that have no real counterpart in our own Solar System \citep{pierrehumbert2013strange}, and pose challenges
to existing theories of planetary formation and evolution. 
Many of the planets are known to have atmospheres or expected to have atmospheres, and
in some cases (though primarily so far only for hot gas giants) there has even 
been progress in characterizing atmospheric composition and some aspects of atmospheric
thermal structure. Understanding the climate dynamics of these novel atmospheres is
an essential part of formulating observational strategies and interpreting observations.
Climate dynamics also enters into determining the long term evolution of atmospheres and
of the surface conditions; the question of habitability occupies a central position in
such studies, but even thoroughly uninhabitable planets can yield important insights.
Exoplanets have opened a pathway to a new adventure in geophysical fluid dynamics,
especially when the fluid dynamics is linked to novel forms of radiative transfer, thermodynamics
and atmospheric chemistry. 
This presents a unique opportunity to imagine atmospheres before they have been observed,
with the virtual assurance that, given the billions of exoplanets in our galaxy alone, 
almost anything we can imagine is likely to be out there somewhere.  

The new era of exoplanet exploration has stimulated work on such novel climate
phenomena as climates of planets in tide-locked or other low order spin-orbit 
configurations where the length of day is comparable to the length of year
\citep{joshi2003TideLock,pierrehumbert2011palette}, climate of transonic rock vapor atmospheres on planets with permanent day-side magma oceans \citep{castan11},
climate of hot Jupiters and hot mini-Neptunes 
\citep{SingClearToCloudyHotJupitersNature2015,HengShowmanAREPS2015}, 
and climate of planets subject to 
the infrared-rich illumination by red dwarf stars \citep{ClimateBook}. This list is far
from exhaustive, and grows by leaps and bounds on a practically monthly basis.  
In this paper, we will focus on some frontiers in geophysical fluid dynamics associated
with atmospheres in which the condensible component is nondilute.

Water vapor is the chief condensible component of the Earth's atmosphere, and
it is dilute in the sense that its molar concentration is never more than 
a few percent, being largest in the warm tropical lower atmosphere.  Many
aspects of conventional terrestrial climate models make heavy use of approximations
appropriate to the dilute limit. For example, while the latent heat released by
condensation is a central driver of the general circulation, the pressure changes
due to mass loss from rainfall have at best minor dynamical consequences.  Similarly,
on the present Earth water vapor has only a minor (though not entirely negligible) 
effect on buoyancy.  There are, however, many planetary circumstances under which
water vapor or other condensibles can become nondilute, as detailed in 
Section \ref{section:WhenNondilute}.  Even in the Solar system, nondilute dynamics is
involved in consideration of the runaway water vapor greenhouse on Venus, the methane
"hydrological" cycle on Titan, and the effects of $\mathrm{CO_2}$ condensation on
both present and ancient Mars.  The hydrological cycle on Earth would edge
into nondilute behavior if the planet were as little as 20C warmer.  Nondilute
atmospheres occur at both the inner edge of the conventional habitable zone where
water vapor is the condensible, and at the outer edge where $\mathrm{CO_2}$ is the
condensible.  They also can be expected in a wide variety of other planetary
circumstances, as detailed in Section  \ref{section:WhenNondilute}. 

For nondilute atmospheres, the mass lost via precipitation leads to significant
horizontal pressure gradients, and thus opens the door to a range of novel
dynamical phenomena.  Transport of energy and angular momentum by precipitation
potentially become major effects in nondilute atmospheres.  Convection in nondilute
atmospheres can be quite different from the familiar dilute case, and poses a challenge
to current approaches to convection parameterization,  not least
because of the challenge of properly enforcing enthalpy conservation in the course
of adjustment. There are boundless possibilities for surprises in the collective behavior
when all these things are linked together and coupled to the rest of the climate system. 

In many of the examples to be discussed in this paper, the condensible component will be taken to be water vapor.
However, the principles illustrated are general and apply equally to any condensible component,
save that the typical temperature at which the atmosphere becomes nondilute will
depend on the thermodynamic properties of the condensible component and both the inventory
and thermodynamic properties of the noncondensible background gas. Similarly we will use
terms like "dry", "moisture" and "relative humidity," which commonly apply to water,
to refer to analogous concepts for other condensibles. 

Any problem involving condensation must confront the difficult issue of microphysics -- loosely speaking the kinetics
governing initial formation of condensate, the subsequent growth of particles by deposition and coagulation, their
transport by precipitation, and their mass loss due to evaporation as they fall through subsaturated layers of the atmosphere.
These phenomena are all of considerable importance. For example, a parcel of atmosphere which is cooled by lifting or
by radiative cooling can become highly supersaturated before condensate forms, in the absence of suitable cloud condensation
nuclei (CCN's) to serve as sites to trigger initial formation of condensate.  The density of CCN's can control the initial
size of condensate particles, which in turn governs how long they take to grow to a size sufficient to cause appreciable
sedimentation. The physical chemistry determining when a particle is a good CCN, and the factors governing the sources
and sinks of CCN's pose formidable challenges even for the well-observed case of dilute condensation of water vapor
on Earth.  There is little doubt that microphysics will prove to be of great importance in governing condensible dynamics
in the nondilute and chemically diverse atmospheres of exoplanets, but we will not attempt to address such issues
in the present work.  As an initial step in the exploration of nondilute dynamics, we will work in a highly 
idealized world in which there is sufficient supply of CCN that condensation always keeps the system from becoming
supersaturated, in which condensate particles grow sufficiently quickly that they can be regarded as instantaneously
removed by precipitation, and in which no mass of precipitate is lost on the way to the planet's surface condensate
reservoir. 

We begin in Section \ref{section:WTG} with a general discussion of the circumstances in which dynamical effects result in weak
horizontal temperature gradients in the absence of condensation effects.  Analytical results on the nature
of nondilute climates are developed in Section \ref{section:Analytic}.  The Exo-FMS general circulation model,
designed to remain valid for nondilute circulations, is described in Section \ref{section:FMSdescription},
and in Section \ref{section:FMSRapidRotSims} we present some preliminary simulation results showing the transition
from dilute to nondilute behavior on rapidly rotating planets.  Our principal findings, and directions for future
research, are summarized in Section \ref{section:discussion}

\section{Preliminaries: The WTG limit}
\label{section:WTG}

One of the key properties of nondilute atmospheres which we will demonstrate is
that they are characterized by weak horizontal temperature gradients.  Therefore,
it is useful to examine other dynamical regimes which produce weak temperature gradients
without relying on nondilute physics.  Loosely speaking, at latitudes where the
effective Coriolis force is weak, temperature gradients are also weak because the pressure
gradients associated with strong temperature gradients cannot be balanced by the Coriolis
forces produced by winds of a magnitude compatible with other dynamical constraints 
(e.g. angular momentum conservation).  The Coriolis parameter giving the local effect of rotation is
$f=2\Omega\sin\phi$ where $\Omega$ is the angular rotation rate of the planet and $\phi$
is the latitude. Thus, the weak gradient behavior always applies near the equator, and
indeed this forms the basis of the standard conceptual model of the Earth's tropical
circulation (see \citet{FalseThermostats2009} for a summary of the ideas and
their history). For a slowly rotating planet, the weak temperature gradient (WTG)
approximation holds globally. This regime is of particular importance for tide-locked
planets about low-mass stars, since tide-locking can occur in long-period orbits in
that case, and in particular within the conventional habitable zone of M-dwarf stars.
The application of WTG in this regime was elaborated in \citet{pierrehumbert2011palette}.
But what determines the value of $\Omega$ for which the rotation rate can be considered slow
enough for the WTG approximation to be valid? A little dimensional analysis helps to resolve this question. 

The basic principles are sufficiently illustrated by examining the zonal momentum
equation written in local Cartesian coordinates near a given latitude.  If we
assume the hydrostatic approximation, the steady state zonal momentum equation takes the form 
\begin{equation}
u\partial_x u + v \partial_y u = - f v -\partial_x g Z'
\label{eqn:MomentumDim}
\end{equation}
where $Z'(x,y,p)$ is the height of the isobaric surface at pressure $p$ with the mean
value over the surface removed and the derivatives are taken at constant pressure.
Now let's nondimensionalize velocities by the characteristic scale $U$, lengths by
scale $\ell$ and $Z'$ by a depth scale $D$. 
Then, in nondimensional form the equation becomes
\begin{equation}
Ro^2 (u\partial_x u + v \partial_y u )= - Ro~v  - \Lambda^2 \partial_x Z'
\label{eqn:MomentumNonDim}
\end{equation}
where
\begin{equation}
Ro\equiv \frac{U}{f\ell} , \Lambda \equiv \frac{\sqrt{gD}}{f\ell}
\end{equation}
and all variables have been replaced by their nondimensionalized counterparts. 
The parameter $\Lambda$ is the ratio of the radius of deformation $L_d \equiv f^{-1}\sqrt{gD}$
to the length scale $\ell$.

From the hydrostatic relation, it can be concluded that the nondimensional $Z'$ characterizes
the magnitude of temperature fluctuations on an isobaric surface. Specifically
\begin{equation}
\frac{\partial Z'}{\partial \ln p} = - \frac{H(\bar{T}(p))}{D} \frac{T'}{\bar{T}(p)} 
\end{equation}
where $\bar{T}$ is the horizontal mean over the isobaric surface, $H$ is the corresponding
local scale height and $T'$ is the deviation from $\bar{T}$. Since $\ln p \approx -z/H $, it follows that  $Z'$ is small
if and only if the the temperature variations (relative to the mean) on an isobaric
surface are small. 

The familiar quasigeostrophic limit for rapid rotators applies for 
$Ro \ll 1$ with $\Lambda^2 = O(Ro)$, in which case the dominant balance is between
the Coriolis force and the pressure gradient force. This situation permits large
geostrophically balanced horizontal temperature gradients.  

Given that $L_d$ represents the radius of influence of
steady-state response to localized heating with vertical scale D, the usual naive criterion for
weak temperature gradients over length scale $\ell$ is $L_d \gg \ell$, i.e. $\Lambda \gg 1$.
This would correspond to globally valid WTG behavior if $\ell$ is taken to be the
planet's radius. However, for slow rotators we typically have $Ro \ge O(1)$, which is
indeed the dynamical definition of a slow rotator.  From Eq. \ref{eqn:MomentumNonDim}
it is evident that $\Lambda \gg 1$ is not sufficient to guarantee WTG behavior; one
in fact also needs $Ro/\Lambda \ll 1$. The typical situation in which WTG behavior
is expected involves large $\Lambda$ with $Ro$ order unity or larger, but $Ro$ must
not be {\it too} large, otherwise strong pressure gradients can be balanced by strong
wind gradients.  Even if $\Lambda \gg 1$, WTG behavior can be broken by the generation of
very strong super-rotating winds, for example. The appearance of $Ro$ in the criterion
for WTG validity makes our scaling theory diagnostic rather than fully predictive, since
$Ro$ is not an externally imposed parameter but rather a property that in principle
should be deduced from the planetary conditions driving the circulation in a complete
scaling theory. Consistency can be checked by computing a circulation based on the WTG approximation,
and then evaluating whether the constraint on $Ro$ is satisfied.  The way this
check plays out will depend on such things as the strength of the differential 
stellar heating of the planet's atmosphere and the radiative damping time, among many
other factors. Because of the lack of a general theory determining the strength
of super-rotating winds, carrying out such a program poses a considerable challenge, and
we shall not attempt to meet it here. One would also need to consider the possibility
that atmospheres might support a WTG state with weak winds, but also an alternate
state with strong winds and temperature gradients. Some attempts at formulating
criteria for WTG validity based entirely on externally imposed parameters are presented 
in \citet{showman2013atmospheric} and \citet{KollAbbotApJ2016}.

Note that $Ro/\Lambda = U/\sqrt{gD}$,which is a Froude number, and that
for deep motions where we can take $D=H$, then the denominator becomes $\sqrt{RT}$, which
is the speed of sound up to an order unity factor.  In this case $Ro/\Lambda$ is in fact
a Mach number, whence we can conclude that $\Lambda \gg 1$ is sufficient to assure
WTG behavior so long as the planetary winds are strongly subsonic.  In the 
cartesian-coordinate momentum equation given in Eq. \ref{eqn:MomentumDim}, violation
of the strongly subsonic condition implies that the nonlinear momentum advection becomes
available to balance pressure gradients; on a sphere, another important nonlinear term
is available -- the cyclostrophic terms $\frac{1}{a}uv\tan\phi$ in the zonal momentum
equation and $\frac{1}{a}u^2\tan\phi$ in the meridional momentum equation, where $a$ is
the radius of the planet. This term has the same scaling as the momentum advection
term for planetary scale circulations where $a$ is the appropriate length scale.  Although
inclusion of the term does not change the conditions for WTG to be valid, it does 
introduce cyclostrophic balance of the pressure gradients due to strong temperature
gradients as a possible state when the conditions for validity of WTG are violated.

To assess the occurence of weak temperature gradients in deep layers on the planetary
scale, we set $\ell$ to the planetary radius $a$ and $D$ to the scale height $H = RT/g$, where
$T$ is a typical temperature of the atmosphere and $R$ is the gas constant corresponding to the
dominant atmospheric composition.  In that case 
$\Lambda = \frac{\sqrt{RT}}{\Omega a}$ and the criterion becomes independant of the surface gravity.
For Earth $\Lambda = .62$, and WTG behavior is confined to the tropics.  GJ1214b with an $\mathrm{H_2O}$
atmosphere at 700K would have $\Lambda = .7$, and would thus be in an Earthlike regime that could
support strong extratropical temperature gradients.  With an $\mathrm{H_2}$ dominated atmosphere,
GJ1214b would have $\Lambda = 2.1$, owing to the greater scale height of the hydrogen atmosphere,
and would exhibit WTG behavior more globally.  In the Solar system Titan has $\Lambda = 15$, and is
deep in the WTG regime. A tide-locked GJ581c with a $\mathrm{CO_2}$ atmosphere at 700K would have 
$\Lambda = 6$ while the more weakly illuminated GJ581d with the same kind of atmosphere at 250K 
would have $\Lambda = 17$ and be in a Titan-like regime. 

The case of Venus illustrates the difficulty of developing a fully predictive
scaling theory for WTG validity. For Venus, $\Lambda \approx 100$ near the surface and
$\Lambda \approx 50$ at the colder temperatures of the high altitude cloud deck but in 
either case one would expect Venus to be deeply into the WTG regime. Indeed, there is
little horizontal temperature gradient near the surface of Venus. However, at the level
of the cloud deck the temperature difference between pole and equator is on the order
of 30K \citep{markiewicz2007VEXWinds}, which is similar to the meridional temperature variation on Earth.  The relatively
large temperature gradient in the Venus cloud deck is made possible by cyclostrophic
balance associated with the strong super-rotating winds in the upper atmosphere
of Venus.  Based on typical wind speeds of 100 m/s \citep{markiewicz2007VEXWinds}, 
$Ro/\Lambda \approx .5 $ , which confirms the importance of the nonlinear cyclostrophic
terms in the momentum equation.  Without an estimate of the Venusian wind speeds -- from
observations, accurate simulations, or theory -- it would not, however, be possible to
anticipate that WTG should break down in the upper atmosphere of Venus. 

Even for slow rotators with weak winds, a large horizontal temperature gradient can be supported if it exists only
in a vertically thin layer, as shallow temperature anomalies create little pressure gradient.  
This remark applies anywhere in the atmosphere, but shallow temperature anomalies typically form
near the surface of a planet with a distinct surface, in cases where thermal coupling between the surface
and the overlying atmosphere is sufficiently weak.  For example, in the tide-locked simulations reported
in \citet{pierrehumbert2011palette}, surface temperature variations are sufficient to allow a pool of open
water having temperatures above 300K near the substellar point even though the rest of the planet is frozen
over and has nightside temperatures as low as 200K. These strong surface temperature variations coexist
with free tropospheric temperatures that are very nearly uniform in the horizontal. 

\section{Analytic Results on nondilute atmospheres}
\label{section:Analytic}

\subsection{Thermodynamic fundamentals}
\label{section:thermo}

\subsubsection{When is nondilute dynamics important?}
\label{section:WhenNondilute}

When a planet has a condensible reservoir at the surface, the partial pressure of the condensible
at the surface will be determined by the Clausius-Clapeyron relation, yielding
$p_c(p_s) = p_{c,sat}(T_s)$, where $p_s$ is the surface pressure, $T_s$ is the surface temperature,
$p_c$ is the condensible
partial pressure and $p_{c,sat}$ is the saturation vapor pressure of the condensible. 
The exponential increase of saturation vapor pressure with temperature will always lead
to nondilute behavior at sufficiently high temperatures. The threshold temperature
increases with the amount of noncondensible gas in the atmosphere, and also depends on the
thermodynamic parameters of the condensible and noncondensible gases.
Suppose that the mass path (mass per square meter of planetary surface)  of the
noncondensible is $\mu_a$, and define the pressure
$p_{a0} \equiv \mu_a g$, where $g$ is the surface gravity. This pressure is
the surface pressure when there is essentially no condensible in the atmosphere,
and it is approximately the partial pressure of noncondensible gas at the surface
when the atmosphere is dilute. Atmospheres over a surface reservoir generally
become nondilute first at the surface where the vapor pressure is greatest; for
atmospheres in which temperature decreases with height, as in convective tropospheres,
the behavior becomes more dilute as altitude increases, owing to the corresponding
decrease in saturation vapor pressure.  Because of the flatness of the moist adiabatic
temperature profile in the strongly nondilute limit (Section \ref{section:Analytic}\ref{section:thermo}\ref{section:SteamLimit}),
strongly nondilute atmospheres can develop deep nondilute layers.

The actual noncondensible partial pressure at the surface deviates from $p_{a0}$ as the
atmosphere becomes nondilute, but comparing the surface vapor pressure to $p_{a0}$
still serves as a useful measure of the threshold of the nondilute regime.
Here, we adopt the criterion that nondilute effects begin to become significant
at a temperature such that the saturation vapor pressure at the surface equals
10\% of $p_{a0}$, corresponding approximately to a 10\% molar concentration of the
condensible; the atmosphere becomes strongly nondilute at somewhat higher
temperatures.   The threshold temperature depends on $p_{a0}$ and the
thermodynamic parameters of the condensible. When $p_{a0}=0$, the atmosphere
is strongly nondilute regardless of the temperature, though at low enough temperatures
the atmosphere may be exceedingly thin and may in fact be too thin to act as a continuum
fluid.  The threshold temperature increases monotonically with $p_{a0}$. 

In Table \ref{table:NondiluteTemperature} we show the threshold temperature for selected
substances making up the surface condensate reservoir, at $p_{a0}=.1$bar and $p_{a0}=1$bar .
Even in the presence of a significant noncondensible background,
nondilute behavior can occur across a wide range of temperatures and background atmospheric
pressures, including cryogenic atmospheres with condensing $\mathrm{N_2}$, rock vapor atmospheres
such as might occur on rocky planets like Kepler10b and 55 Cancri Ae in close orbits about their stars, and all points in between. 
The rock vapor case is represented by carbon, iron, enstatite and (on the low end for substances considered "rocks or minerals" on Earth) sulfur,
but there are many other possibilities in a similar range, including sodium and SiO. 
The threshold temperature for nondilute behavior, if based on molar concentration, depends on $p_{a0}$ but not
on the composition of the noncondensible gas, so long as the background gas remains noncondensing over the
range of temperatures under consideration. For example, around 60K one can have condensing $\mathrm{N_2}$ in a background
of noncondensing $\mathrm{H_2}$ and $\mathrm{He}$, or at temperatures around 180K condensing $\mathrm{CO_2}$ in a background of noncondensing
$\mathrm{H_2}$, $\mathrm{He}$ and $\mathrm{N_2}$. The relative molecular weights of the condensible and noncondensible do, however, affect
the mass fraction of the noncondensible at a given molar concentration, and many physical effects (e.g. surface pressure changes due
to precipitation) are more directly tied to mass fraction than molar fraction. Viewed in terms of mass fraction, condensing $\mathrm{CO_2}$
in noncondensing $\mathrm{H_2}$ would be much more nondilute at the $\mathrm{CO_2}$ threshold temperatures given in Table \ref{table:NondiluteTemperature} than would
condensing $\mathrm{H_2O}$ in noncondensing $\mathrm{CO_2}$ at the $\mathrm{H_2O}$ threshold temperatures. 


\begin{table}[!h]
\caption{Threshold temperature $T_{nd}$ above which behavior begins to become nondilute
at the surface. $p_{a0}$ is the surface pressure the noncondensible gas would have
in the absence of any condensible, and is a measure of the mass of noncondensible
in the atmosphere. The system is considered to be on the threshold of nondilute
behavior when the molar concentration of condensible at the surface equals 10\%. $p_{a0}$
is given in bars, and $T_{nd}$ in kelvins.Data for
Fe and $\mathrm{MgSiO_3}$ are from \citet{cooper2003modeling,ackerman2001precipitating,BarshayLewis1976}.}
\label{table:NondiluteTemperature}
\begin{tabular}{lll}
\hline
Condensible & $T_{nd}$, $p_{a0}=.1$ bar &$T_{nd}$, $p_{a0}=1$ bar\\
\hline
$\mathrm{C}$           & 3293 & 4002 \\
$\mathrm{Fe}$          & 2566  & 3052   \\
$\mathrm{MgSiO_3}$ (enstatite)    & 1957  & 2120 \\
$\mathrm{S}$           & 446  & 550 \\
$\mathrm{H_2O}$        & 280  & 318 \\
$\mathrm{NH_3}$        & 180  & 201  \\
$\mathrm{CO_2}$        & 151  & 170 \\
$\mathrm{CH_4}$        & 76   & 90 \\
$\mathrm{N_2}$         & 53   & 62 \\\hline
\end{tabular}
\end{table}

A planet with a condensate reservoir at the surface will exhibit nondilute dynamics
when the surface temperature exceeds the threshold corresponding to the composition
of the condensible and the amount of noncondensible background gas in the atmosphere.
For this situation to apply, there must be enough total inventory of condensible substance 
that there is some left in the surface condensed reservoir at temperatures high enough to
make the atmosphere nondilute.  The nondilute layer in this case will extend upwards from the surface. 

This situation occurs over a liquid water ocean (given sufficient total water inventory) as the inner edge
of the habitable zone -- defined by the water vapor runaway greenhouse -- is approached from the habitable side.
It also occurs over a liquid $\mathrm{CO_2}$ ocean or a $\mathrm{CO_2}$  glacier somewhat past the outer edge
of the conventional habitable zone, near the point where a $\mathrm{CO_2}$ runaway occurs. This would typically occur
on planets illuminated somewhat more faintly than current Mars, 
but which (unlike current Mars) are sufficiently carbonate-rich and which have sufficiently active tectonics to support substantial
outgassing of $\mathrm{CO_2}$ from the interior. 
Nondilute dynamics also generally prevails during the window of time when a planet is actually
undergoing a runaway, but before it has lost its surface condensate reservoir to the atmosphere.
There is only a narrow window of time of some thousands of years during which this situation
persists, but it is still important to understand the dynamics in this regime, since
clouds or subsaturation could in principle act to arrest the runaway.   

Given a sufficient inventory of condensible, nondilute dynamics will generically occur on any planet 
whose instellation is below the threshold for runaway with regard to its condensible substance, but whose atmosphere
also contains enough of a powerful noncondensing greenhouse gas to bring the surface temperature to the threshold
where nondilute behavior sets in.  This occurs because a planet cannot enter a runaway state if the instellation
is below a certain threshold (called by some the K-I limit \citep{ClimateBook}), no matter how high the surface
temperature is made. The noncondensing greenhouse gas must be a potent one, otherwise the increase in temperature it
causes as more is added to the atmosphere can be outstripped by the increase in the threshold temperature for nondiluteness,
associated with the increase in $p_{a0}$.  
The collisional opacity of noncondensing $\mathrm{N_2}$ on Titan brings
methane to the brink of nondilute behavior \citep{lorenz1999TitanStability}. The Earth's atmosphere would start to become nondilute if the $\mathrm{CO_2}$
content increased to the point that surface temperatures exceeded 320K, a situation that could have occurred during
a post-Snowball hothouse \citep{pierrehumbert2011AREPSNeoprot}, or on a watery world with insufficiently strong silicate weathering.
On planets with an $\mathrm{H_2}$ supported greenhouse \citep{pierrehumbert2011hydrogen}, nondilute dynamics could occur with regard
to many of the condensibles listed in Table \ref{table:NondiluteTemperature}. For example, consider a planet with a $\mathrm{CO_2}$
glacier at the surface, in an orbit such that the absorbed stellar radiation is 17 $\mathrm{W/m^2}$ averaged over the surface.
If the only source of atmosphere were sublimation of $\mathrm{CO_2}$ from the glacier, the surface temperature would be about
132 K and the planet would have a thin nondilute pure $\mathrm{CO_2}$ atmosphere with surface pressure of 45 Pa.  Using the same 
radiative transfer calculation as described in \citet{pierrehumbert2011hydrogen}, giving the planet a 1 bar $\mathrm{H_2}$ background atmosphere
would raise the surface temperature to 200 K and the surface $\mathrm{CO_2}$ partial pressure to 1.6 bar, resulting in
a thick strongly nondilute atmosphere. (This calculation assumes a surface gravity of 17 $\mathrm{m/s^2}$, which is typical for
a rocky Super-Earth). If the initial atmosphere, instead of being pure $\mathrm{CO_2}$ had also contained .1 bar ($10^4$Pa) of noncondensing
$\mathrm{N_2}$ (which is not enough to add significant opacity), then the addition of the $\mathrm{H_2}$ would have transformed
the dynamics from strongly dilute to strongly nondilute.  Similar $\mathrm{H_2}$ supported nondilute regimes can easily be found for 
$\mathrm{NH_3}$, $\mathrm{CH_4}$ and $\mathrm{N_2}$. Where $\mathrm{N_2}$ is present, either as a condensible or as a noncondensible
background gas, the $\mathrm{N_2-H_2}$ collisional continuum can provide an additional source of opacity \citep{wordsworth13}. 

The cold low-pressure atmospheres that form by sublimation from icy bodies form another important class of atmospheres where nondilute
dynamics is important.  This regime is familiar from the case of Triton in the Solar system, which has an approximately 2 Pa global $\mathrm{N_2}$
atmosphere, which is already thick enough to  exhibit interesting dynamics as a continuum fluid \citep{ingersoll1990Triton}. 
Similar atmospheres would occur for planets enveloped in a global $\mathrm{CO_2}$ ice glacier, though we have no examples of such
a planet in the Solar system; while there are no exoplanet observations that confirm the existence of such a planet,
this sort of behavior is expected on any planet in a sufficiently distant orbit whose tectonics and composition support substantial
outgassing of $\mathrm{CO_2}$. Water ice can also feed a significant pure water vapor atmosphere. In the Solar System, Europa
is too cold to do this, as the vapor pressure of water at the subsolar point is on the order of $10^{-12}$ Pa . However, for a 
Snowball Earth with a surface temperature of 265K and no other atmosphere besides that sublimated from the water ice surface,
the surface pressure would be 305 Pa, or half that of the present Mars.  Such thin, pure water-vapor atmospheres have been studied
in connection with loss of water to space and formation of abiotic oxygen signatures \citep{wordsworth2014abiotic}. 
In any of these cases, nondilute behavior can survive the addition of a background noncondensible gas (e.g. $\mathrm{H_2}$ in
the condensing $\mathrm{N_2}$ case, or $\mathrm{N_2}$ in the condensing $\mathrm{CO_2}$ or $\mathrm{H_2O}$ cases) if the surface
temperature is increased sufficiently owing to some combination of higher instellation or addition of a noncondensible greenhouse gas.

Another class of low-pressure pure-condensible atmosphere forms when the rate of heat loss from the condensible gas is so great that 
the gas does not travel very far from its source before most of it condenses back out onto the planetary surface.  In such cases,
the strong pressure gradients typically drive supersonic flow.  Such atmospheres were first studied by Ingersoll \citep{ingersoll85}
in connection with local $\mathrm{SO_2}$ atmospheres on Io, which can be sourced either from sublimation of $\mathrm{SO_2}$ frost at the
subsolar point or from transient volcanic outgassing.  Ironically, similar behavior has been proposed for extremely hot rock vapor
atmospheres, such as probably occur on Kepler 10b \citep{castan11}. Many volatiles are possible in rock-vapor atmospheres, but for
55 Cancri Ae we highlight carbon as an intriguing possibility because of conjectures that this planet might be a carbon-rich body with
a graphite surface \citep{madhusudhanCarbon2012}. 
The dynamics of local supersonic or transonic atmospheres are very different from the hydrostatic circulations treated in general
circulation models based on the primitive equations.  This contrasts with the thin $\mathrm{N_2}$ atmosphere of Triton,
which is global and largely subsonic.  A general theory determining when one gets local vs global atmospheres has not yet been formulated. 
Additionally, the introduction of a sufficient inventory of noncondensible background gas would presumably even out surface pressure variations
and result in a subsonic hydrostatic circulation regime, but the way the circulation makes the transition from supersonic and local
to subsonic and global as the noncondensible inventory increases has not yet been systematically explored. 



Nondilute behavior can occur aloft without a surface reservoir of condensible; this
would be the typical situation in a post-runaway atmosphere in which the surface ocean
or glacier has been depleted so that the lower atmosphere becomes hot and unsaturated.
Nondilute condensing dynamics aloft does not even require the presence of a distinct
planetary surface at all. This is important because of the prevalence of fluid
super-Earths such as GJ1214b. One of the conjectured compositions for GJ1214b is
nearly pure water vapor, and while GJ1214b itself is too hot for water vapor to condense
to any appreciable degree within the atmosphere, a similar planet in a somewhat more
distant orbit would inevitably have a nondilute condensing layer. Under some circumstances,
to be outlined below, the layer could remain nondilute even in the face of admixture
of some noncondensible background gas such as $\mathrm{H_2}$ or $\mathrm{N_2}$

The lifted condensation level is determined by the relation
\begin{equation}
p_{c,sat}(T_\mathrm{dry}(p_\mathrm{lcl})) = \eta p
\end{equation}
where $T_\mathrm{dry}(p)$ is the dry adiabat corresponding to the composition of the atmosphere and
$\eta$ is the (constant) molar concentration of the condensible in the noncondensing layer, which
is assumed to be well-mixed. This relation proceeds directly from the definition of $\eta$.
It follows that the condensing layer will be nondilute when $\eta$ exceeds a specified threshold
value (10\% in the criterion we have been using). 

\begin{figure}[h]
\centering\includegraphics[width=2.5in]{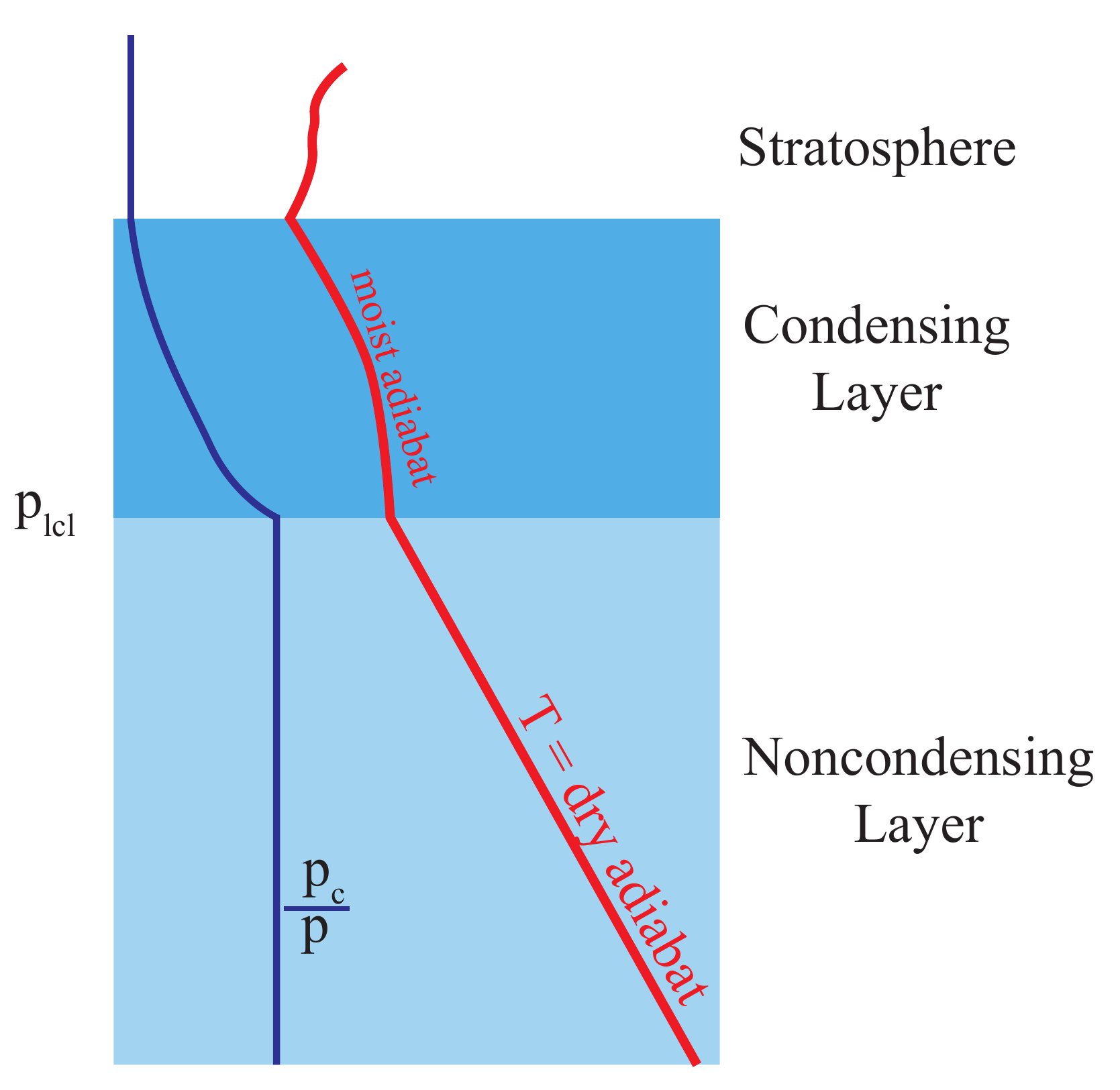}
\caption{Configuration of an atmosphere with a deep non-condensing layer and a condensing layer aloft. This is typical of the situation in which
the condensate reservoir at the surface (if there is one) has been depleted, or on gaseous planets with no effective surface at all. 
$p_\mathrm{lcl}$ is the lifted condensation level, and $\eta = p_c/p$ is the molar concentration of the condensible substance.}
\label{fig:CondensingLayer}
\end{figure}

The condensing layer in Fig. \ref{fig:CondensingLayer} can be eliminated if the temperature of the upper atmosphere -- presumed to
be in radiative equilibrium -- is sufficiently hot that the atmosphere never intersects the condensation threshold.  This would 
typically happen as a result of strong absorption of shortwave stellar radiation, but even if the upper atmosphere is in pure
infrared radiative equilibrium, the radiative equilibrium temperature of the upper atmosphere can be approximately uniform in
height in the optically thin portion of atmospheres with broadband absorbers \citep[Chapter~4]{ClimateBook}, and it is possible for this temperature to be above the condensation threshold. 
 
Fig. \ref{fig:CondensingLayer} assumes a convective layer below the condensing layer, but the situation does not change greatly
if the deep atmosphere is radiative instead, since pure infrared radiative equilibrium for an optically thick atmosphere (even one
that is only optically thick in a small portion of the spectrum) also has temperature decreasing with altitude.  On the other hand,
if the atmospheric absorption of shortwave stellar radiation is so strong that essentially all of the instellation is depleted
in the upper atmosphere, and if in addition the atmosphere does not need to carry any significant heat flow from the 
interior of the planet,  then the deep atmosphere is isothermal, and the temperature of the deep isothermal layer 
determines whether the atmosphere has a very deep condensing layer or no condensing layer at all. Note, however, that when
the deep atmosphere is very optically thick in the infrared, it takes very little heat flux (and very little deep penetration
of stellar radiation) to support a convective layer or a radiative layer with a strong vertical temperature gradient. This
is the case, for example, in the atmosphere of Venus, for which most of the atmosphere remains near the $\mathrm{CO_2}$ dry
adiabat despite the feeble penetration of solar radiation to the surface. 

Near the outer edge of the conventional habitable zone, $\mathrm{CO_2}$ condensation becomes prevalent on planets which can maintain
a substantial $\mathrm{CO_2}$ inventory in their outer envelope, for example by volcanic outgassing from decomposition of carbonates in the 
deep interior of the planet. On the warm side of the outer limit, one would get nondilute layers which do not reach the surface, as in
Fig. \ref{fig:CondensingLayer}. This situation would be expected in the $\mathrm{CO_2}$ rich atmospheres of a warm, wet Early Mars,
or on a Super-Earth like GJ581d \citep{wordsworth2011gliese} if made habitable by a massive $\mathrm{CO_2}$ atmosphere.  
If somewhat beyond the cold edge
of the conventional habitable zone, such planets would instead develop $\mathrm{CO_2}$ oceans or glaciers at the surface, with
a nondilute layer extending all the way to the condensible reservoir at the ground. In either case, nondiluteness would survive
in the face of the addition of a moderate amount of noncondensible background gas such as $\mathrm{N_2}$ or $\mathrm{H_2}$.


\subsubsection{The "pure steam" limit}
\label{section:SteamLimit}

Consider a substance which can undergo a phase transition from a gaseous (also 
known as vapor) phase to a condensed phase, which may be liquid or solid. Over a 
temperature range within which the latent heat $L$ of the phase transition can be
considered constant, the Clausius-Clapeyron relation determining the phase boundary
in pressure-temperature space implies

\begin{equation}
p_{c,sat} = p_1 \exp\Big(-\frac{L}{R_c} (\frac{1}{T} - \frac{1}{T_1})\Big)
\label{eqn:ClausiusSimp}
\end{equation}
where $p_{c,sat}$ is the partial pressure of the condensible substance, $R_c$ is the
ideal gas constant for that substance and $(p_1,T_1)$ is any point on the phase
boundary.  $p_{c,sat}$ is referred to as the saturation vapor pressure, or sometimes
simply vapor pressure. The derivation of Eqn. \ref{eqn:ClausiusSimp} also assumes the
ideal gas equation of state and that the density of the condensate is much greater than the
density of the vapor phase. The latter assumption breaks down near the critical point, as
does the assumption of constant latent heat, since $L \rightarrow 0$ as the critical point
is approached. Thus, Eq. \ref{eqn:ClausiusSimp} cannot be used near the critical point.

If the vapor phase is in contact with the condensed phase and
comes into thermodynamic equilibrium with it, the partial pressure of the gas
phase is $p_{c,sat}$.  This contact could take the form of an interface with a macroscopic
reservoir of condensate, such as an ocean or glacier, or it could take the form
of contact with condensate interior to the atmosphere, such as a cloud, snow or rain
particle.  In saturated conditions, the partial pressure of the condensible, $p_c$ is
equal to $p_{c,sat}$, so in the remainder of the exposition we will drop the extra subscripts
when dealing with saturated conditions, when there is no risk of ambiguity.  

Eqn. \ref{eqn:ClausiusSimp} can be solved for $T$ in terms of $p_c$ to yield
\begin{equation}
T_\mathrm{sat}(p_c) = \frac{T_1}{1 - \Gamma \ln\frac{p}{p_1}}, \Gamma \equiv \frac{R_cT_1}{L}
\label{eqn:dewpoint}
\end{equation}
where we have introduced the notation $T_\mathrm{sat}$ to indicate the temperature
obtained by solving Eqn. \ref{eqn:ClausiusSimp} for temperature. 
This is the dew point or frost point formula (according to the nature of
the condensed phase), which gives the temperature below which the condensible
substance will form condensate, if given the chance to reach thermodynamic
equilibrium. For an atmospheric temperature $T < T_{sat}(p_c)$,
the atmosphere is supersaturated. An atmosphere that is sufficiently supersaturated will spontaneously form condensate, increasing temperature via latent heat release
and reducing $p_c$ by conversion of vapor to condensed phase until $p_c$ is reduced to
the saturation vapor pressure corresponding to the new, warmer temperature.  
In the presence of suitable condensation nuclei,
little, if any, supersaturation is required to initiate the formation of condensate.
On the other hand, if the atmospheric temperature is greater than the dew/frost point 
temperature, the atmosphere is subsaturated and condensate does not form.  

If the condensible substance $c$ is the {\it only} gas present in the atmosphere, then
the partial pressure $p_c$ is also the total atmospheric pressure $p$, so the temperature
profile of a saturated atmosphere is simply $T_{sat}(p)$.
This is the saturated moist adiabat for a single-component atmosphere, e.g. 
a pure water vapor atmosphere \citep[Chapter~2]{ClimateBook}. Earth's atmosphere,
in contrast, contains a mix of condensible water vapor with other gases that do
not condense over the range of temperatures encountered in the atmosphere. Generalizing
from the water vapor case, we shall refer to single-component condensible atmospheres as 
"steam atmospheres."  The definition of "condensible" is relative to a planet's climate,
since any gas can be made noncondensible if the atmosphere is everywhere hot enough, and
any gas can condense if made cold enough. The present $\mathrm{CO_2}$ dominated atmosphere of Mars would thus be considered a close approximation to a steam atmosphere in our definition,
as $\mathrm{CO_2}$ condenses in the upper troposphere and in a deeper layer near the winter
poles. 

Because of the exponential growth of vapor pressure with
temperature, an atmosphere with a noncondensible component will generally converge onto
the steam atmosphere limit at high temperatures if the atmosphere is in contact
with a sufficiently large reservoir of condensate, because the mass of condensible in the
atmosphere increases whereas the mass of the noncondensible gases remains fixed. 

For steam atmospheres, the saturated moist adiabat giving temperature as a function
of pressure has no free parameters, as the only constants in Eq. \ref{eqn:dewpoint} describe
thermodynamic properties of the gas. For mixed atmospheres containing a noncondensible
component, the saturated moist adiabat contains a free parameter, namely the mass concentration
$q_a$ of the noncondensible gas at some given pressure level. This allows the temperature
at a reference pressure level $p_0$ to be chosen freely, subject to the constraint that
$T(p_0)$ be less than $T_{sat}(p_0)$, at which temperature $q_a = 0$.  As $T\rightarrow 0$, $q_a \rightarrow 1$. Once $q_a(p_0)$ is known, the saturated moist adiabat is uniquely determined.
For mixed atmospheres the saturated moist adiabat is thus a one-parameter family of curves. 
$q_a$ is constrained by the total mass of noncondensible gas in the atmosphere, which must
remain fixed, but this is a global constraint obtained by integrating over the entire
atmosphere, and cannot be applied locally. 
In contrast,  within any saturated layer of a pure steam atmosphere, the temperature is known
once the pressure is known without any need to supply additional information. 

$\Gamma$, shown for selected gases in Table \ref{table:Gamma} is typically a small number. According to Eq. \ref{eqn:dewpoint}, 
the temperature on the saturated pure steam adiabat varies only from $2T_1$ to $.5T_1$ between pressures  $p_1 \exp(.5/\Gamma)$ and
$p_1\exp(-.5/\Gamma)$ where ($p_1$,$T_1$) is the point on the phase boundary where $\Gamma$ is computed.  This estimate is based on an assumption
of nearly constant latent heat, as is Eq. \ref{eqn:dewpoint}, and so must be modified if the high-pressure and temperature 
end of the range approaches the critical point, where latent heat falls to zero. Even when the pressure is outside the
range of relatively flat temperatures, the temperature approaches zero only gradually with height (like $1/\ln(p/p_1)$, which is approximately\ like $1/z$ in altitude). 

\begin{table}
\caption{Values of $\Gamma \equiv R_c T_1/L$ for selected gases. Based on triple-point data, except for $\mathrm{MgSiO_3}$ (enstatite)
where values are taken at 2310K, where the vapor pressure is 1 bar.  For C and S a mono-atomic vapor is assumed. Data for
Fe and $\mathrm{MgSiO_3}$ are from \citet{cooper2003modeling,ackerman2001precipitating,BarshayLewis1976}.}
\label{table:Gamma}
\begin{tabular}{clllllllll}
\hline
Condensible & $\mathrm{C}$ & $\mathrm{Fe}$ & $\mathrm{MgSiO_3}$ & $\mathrm{S}$ &  $\mathrm{H_2O}$ & $\mathrm{NH_3}$ & $\mathrm{CO_2}$ & $\mathrm{CH_4}$ & $\mathrm{N_2}$ \\
\hline
$\Gamma$    & .112  &.048 &  .039   & .068         & .0506            & .0576           & .103            & .0879           & .0860 \\ \hline
\end{tabular}
\end{table}

\begin{figure}[h]
\centering\includegraphics[width=2.5in]{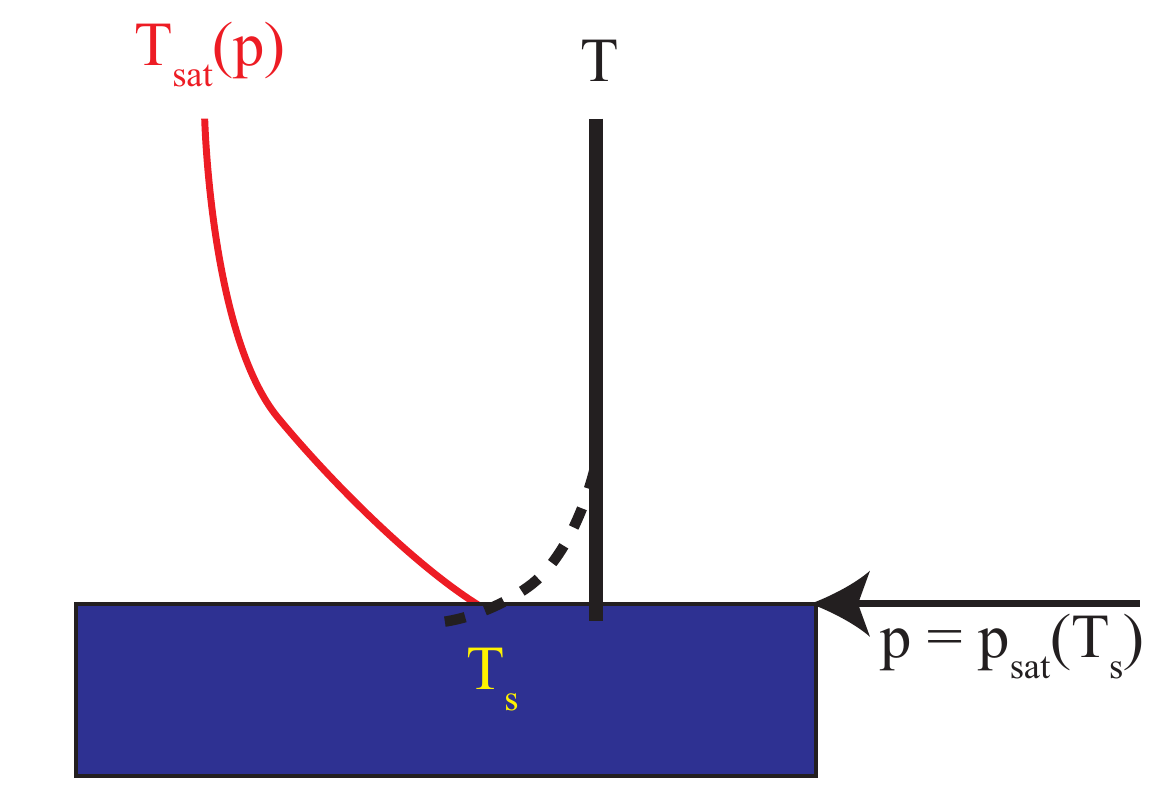}
\caption{Dashed line schematically shows adjustment to equilibrium of an initially isothermal subsaturated pure steam atmosphere over
a condensate reservoir that is colder than the atmosphere, assuming the hydrostatic surface pressure of the atmosphere is continuous with the saturation
vapor pressure corresponding to the ocean surface temperature.
It is assumed the heat capacity of the surface ocean is enough to keep its temperature essentially constant in the course of the adjustment. Heat is transferred from the
atmosphere to surface reservoir in the course of this process, but no mass exchange
takes place.}
\label{fig:SurfExchangeT}
\end{figure}

For dilute atmospheres, evaporation from the surface reservoir occurs whenever the 
overlying atmosphere is subsaturated. However, the lower boundary condition for 
nondilute atmospheres works very differently, particularly as the pure steam
limit is approached. To illustrate this point we'll consider a thought experiment
involving an initially subsaturated atmosphere in contact with the surface condensed
phase reservoir.  Suppose the surface pressure in hydrostatic balance is $p_s$,
as determined by the mass of the atmosphere. Then, subsaturation at the surface is
defined by $T(p_s)>T_{sat}(p_s)$. Let the temperature of the condensed
reservoir be $T_s$, and suppose that $T_s = T_{sat}(p_s)$, which is cooler than
the overlying air temperature. In this case the vapor pressure of the layer of
vapor that forms immediately over the surface is identical to the pressure of
the overlying air, $p_s$, and there is no unbalanced vertical pressure gradient
to drive a bulk mass flow. However, since $T(p_s)>T_s$, radiative, diffusive or
turbulent heat transfer will cause the lower portions of the atmosphere to cool
until $T(p_s)$ approaches $T_s$, forming a stable thermal boundary layer, within which
$T(p)$ increases with height. This situation is illustrated schematically in Fig. \ref{fig:SurfExchangeT}
for the case of an initially isothermal atmosphere. 
The entire layer will still be subsaturated, reaching
marginal saturation just at the interface with the surface reservoir, so there will
be no mass exchange with the reservoir in this situation.  This contrasts with
the more familiar situation in the dilute case, where one can have a subsaturated 
atmosphere with $T(p_s) = T_s$, requiring no heat exchange but allowing evaporation.

\begin{figure}[h]
\centering\includegraphics[width=5in]{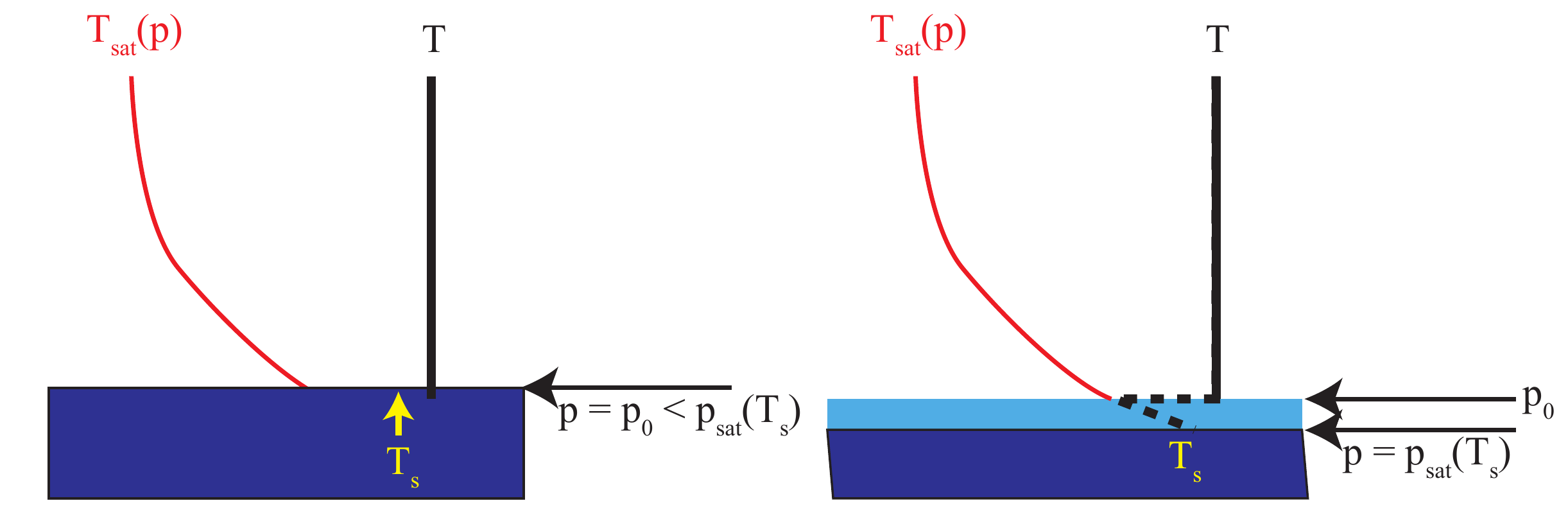}
\caption{Mass and temperature adjustment of an initially subsaturated isothermal atmosphere in contact with a condensate reservoir at the surface. In contrast to the case depicted
in Fig. \ref{fig:SurfExchangeT}, in this case it is assumed that the hydrostatic pressure
at the bottom of the atmosphere is less than the saturation vapor pressure corresponding
to the surface temperature. The left hand panel shows the initial state.  The pressure discontinuity is eliminated by transfer of mass
from the surface ocean to the atmosphere, which adds a layer of new atmosphere near the
surface, thus increasing the surface pressure at the bottom of the atmosphere. New mass 
is added at a temperature following the pure steam adiabat, but the temperature at
altitudes above the pressure level $p_0$ is not affected, leading to an adjusted profile
with an inversion, as depicted in the right hand panel. Subsequent vertical heat
transfer could erode the inversion, extending the subsaturated layer toward the surface. }
\label{fig:SurfExchangeMass}
\end{figure}

Mass exchange between the surface reservoir and the atmosphere is in fact
determined by the vapor pressure $p_\mathrm{sat}(T_s)$ relative to the hydrostatically
balanced pressure $p_s$ given by the mass of the atmosphere. In a pure steam
atmosphere, when $p_\mathrm{sat}(T_s)> p_s$, a large unbalanced pressure gradient will
develop near the surface, driving a rapid flow of mass from the condensate reservoir
into the atmosphere, and correspondingly cooling the surface by evaporation. The increase
in $p_s$ through the addition of mass, and the decrease in $T_s$ (and hence $p_\mathrm{sat}(T_s)$)
through evaporation will continue until the system comes into equilibrium with the
pressure gradient eliminated, as depicted in Fig. \ref{fig:SurfExchangeMass} for an idealized
thought experiment involving an initially isothermal atmosphere. 
Conversely, if $p_\mathrm{sat}(T_s) < p_s$ the condensible vapor
in contact with the surface will condense, releasing latent heat in the atmosphere, and
the pressure gradient will rapidly carry more mass to the surface until equilibrium
is restored. 

In fact, even when the atmosphere consists of a mixture of a condensible and noncondensible
component, the usual bulk formula for evaporative mass exchange between surface
and atmosphere still implies that the exchange rate is proportional to the 
difference between $p_\mathrm{sat}(T_s)$ and the {\it partial} pressure of the condensible
in the overlying atmosphere \citep[Eq. 6.13]{ClimateBook}, which reduces to
the above result in the pure steam limit. The only real difference between the
condensible-rich case and the dilute case is that in the former the difference
between $p_\mathrm{sat}(T_s)$ and the condensible partial pressure is dynamically significant,
and directly drives the flow which acts to restore equilibrium.

\subsubsection{Basic nature of convection in nondilute atmospheres}
\label{section:Convection}

For a noncondensing (i.e. dry) atmosphere, a dry adiabat is established in the 
troposphere through buoyancy-driven mixing. With a condensible component, the
situation is somewhat more complicated since condensation and latent heat release
occur only for upward motions, but in the dilute limit the effect of latent heating
in the upward plumes is communicated to the downward branches through compressive
heating of the noncondensible gas which dominates the mixture, and loosely speaking
the buoyancy-driven convection still mixes the profile towards a moist adiabat. In
either case, buoyancy can be generated either by radiative heating and cooling in
the interior of the atmosphere (when the radiative equilibrium is unstable to convection)
or at the planet's surface (if it has one) through absorption of stellar radiation there.
Convection can also be sustained by heat flux from the planet's interior, especially in 
the case of planets with a deep fluid layer such as Jupiter or Neptune in our own solar system,
or fluid super-Earths such as GJ1214b. 
For nondilute atmospheres, convection works very differently as the steam atmosphere
limit is approached, in fact so differently it is unclear whether it should be called
convection at all. 

For an atmosphere with a noncondensible component, buoyancy can be generated because
the moist adiabat has a free parameter; for example, stellar heating near the planet's
surface can set the boundary layer air to a {\it hotter} adiabat than that characterizing
the free troposphere, so that once a parcel from the boundary layer is lifted to an altitude
where condensation begins, further lifting will follow a saturated moist adiabat which will
eventually render it buoyant with respect to its surroundings. As the atmosphere becomes
increasingly nondilute and all adiabats collapse onto the pure-steam adiabat, 
the ability to generate buoyancy disappears. In fact, the buoyancy referred to in the
preceding discussion is just the thermal buoyancy corresponding to a situation in 
which condensate is instantaneously removed; the mass of retained condensate reduces
buoyancy, and thus exerts a stabilizing effect which may be particulary important as the
pure steam limit is approached. 

As an illustration of the novel features of the strongly nondilute case, consider a
pure steam atmosphere above a condensate reservoir, and neglect all radiative heating
and cooling of the atmosphere for the moment. Assume the pressure to be continuous with the
surface vapor pressure.  Barring supersaturation, the initial
atmospheric temperature profile satisfies $T(p) \ge T_\mathrm{sat}(p)$, and since 
lifted air parcels follow $T_\mathrm{sat}(p)$ they will be negatively buoyant
with regard to the environment wherever the environment is subsaturated, even if
the lifted parcels immediately lose their condensate. 

\begin{figure}[h]
\centering\includegraphics[width=5in]{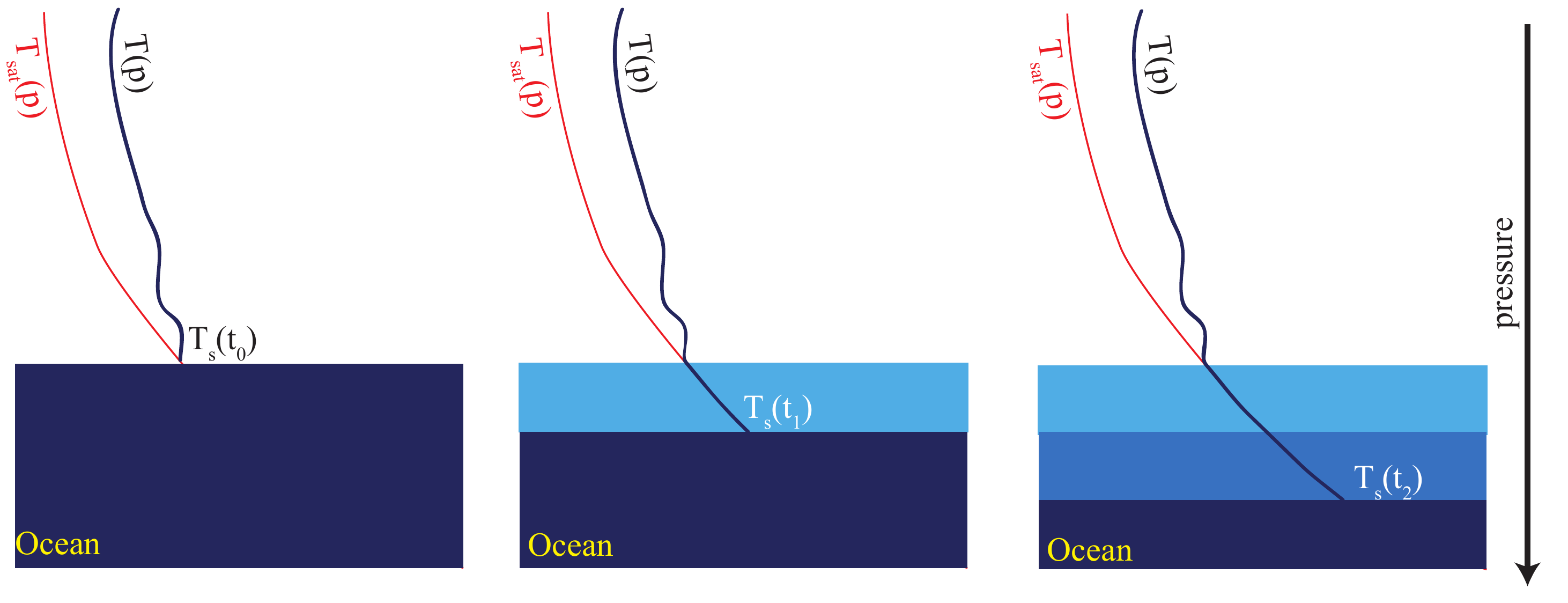}
\caption{Sketch of evolution of a pure steam atmosphere in contact with a surface condensate
reservoir whose surface temperature is increasing with time. The light and dark blue shaded
regions indicate layers of vapor added to the atmosphere by evaporation from the condensate
reservoir.}
\label{fig:SteamAtmWithWarmingSurf}
\end{figure}

Suppose now that absorption of stellar radiation
at the ocean surface causes the surface temperature to increase with time, so $T_s = T_s(t)$
is monotonically increasing. In a dilute atmosphere, this would create buoyancy and eventually deep convection.  The corresponding surface pressure $p_{sat}(T_s(t))$ also
increases monotonically with time, which via the hydrostatic relation corresponds to adding
new mass at the bottom of the atmosphere, as illustrated in Fig. \ref{fig:SteamAtmWithWarmingSurf}. 
In a pure steam atmosphere this
process proceeds without ever creating any buoyancy. 

In pressure coordinates there is no vertical motion associated with this process.
Each new layer of atmosphere is slid into place at its own appropriate pressure
as the boundary between atmosphere and ocean moves to higher pressures.  Seen in
altitude coordinates, there is vertical motion, since the existing atmosphere must
be displaced upward to make room for the new atmosphere that is added at the
bottom, given that the vapor added has lower density and occupies more volume than
the condensed phase removed from the ocean. 

If one allows for net radiative cooling of the atmosphere, then the system can 
reach an equilibrium in which mass loss due to radiatively-driven condensation
aloft equals the mass added at the bottom of the atmosphere by evaporation. 
In this case there is a net vertical motion in pressure coordinates, which gently
lifts the moisture from the surface where it is added to the atmosphere to condensing
levels aloft where it is removed, in a form of laminar moisture elevator.  
This motion is quantified in \ref{section:MassBalance}.

In the general nondilute case that hasn't converged on the pure steam limit, there is
some buoyancy generation and consequent convection, but the role of buoyant plumes is
much less than it is in the familiar dilute case. 
Parameterization of convection for general nondilute atmospheres requires consideration
of a number of additional factors that can be ignored in the dilute limit.
Conservation of enthalpy in the course of mixing by convection needs to take into
account the enthalpy, kinetic and potential energy carried by precipitation, and the
portion of this energy associated with precipitation reaching the surface must
be added to the surface energy budget. These effects are still somewhat significant even 
for the present Earth's relatively dilute climate, and their neglect can lead to 
spurious imbalances in the top-of-atmosphere energy budget. 
Also, loss of mass by precipitation causes
a reduction in pressure in lower levels, which has both dynamical effects through horizontal
pressure gradients and thermodynamic expects through adiabatic expansion to accomodate
the pressure reduction. In \citet{FengNondiluteColumnModel} we have developed
a simple convective adjustment scheme that can deal with the full range from
dilute to steam atmosphere limits in an energetically consistent way, and used
it to further discuss the behavior of nondilute convection.

\subsubsection{Mass balance and Hadley circulations for nondilute atmospheres}
\label{section:MassBalance}

In primitive equation models, vertical velocity is diagnosed from the mass continuity equation.  In the nondilute case, the continuity equation needs be modified taking the mass loss due to precipitation into account. Let us consider the idealized case in which precipitation that forms is removed instantaneously to the surface (or to some deep layer of the atmosphere which we are not explicitly modeling). We may nonetheless allow for the possibility that some of the 
precipitation evaporates on its way down.   Assuming the hydrostatic approximation, 
\begin{equation}
 \partial_p\omega + \nabla \cdot (\mathbf{v}) = -(P-E)
\label{eqn:MassContinuity}
\end{equation}
where $\mathbf{v}$ is the horizontal velocity, $\omega$ is the vertical velocity of the gas phase in pressure coordinates,  $P$ is the local rate of mass precipitation (positive being mass loss) and $E$ the local evaporation rate, both per unit mass of the atmosphere. The idealization
of instantaneous removal enters in an important way in this form of the continuity 
equation, since incorporating the effects of retained condensate would require one
to write separate continuity equations for the gas phase and condensed phase; Eq. \ref{eqn:MassContinuity} would remain valid with $p$ interpreted as total pressure, but
$\omega$ would represent only the rate of change in the hydrostatic pressure due to the gas phase and to be useful would have to be supplemented by a transport equation for the
condensed phase. 

Consider the idealized case of an ascending saturated plume in the pure steam limit,
assuming that precipitation forms rapidly enough to prevent supersaturation, and
that there is no re-evaporation on the way down (so $E=0$).  
$P$ can be computed by determining the amount of latent heat that must be released to warm a parcel
lifted along the noncondensing adiabat to the saturation temperature.  By making use
of the Clausius Claperyon relation and the formula for the dry adiabat, we find
\begin{equation}
\partial_p\omega  \approx -P =  \frac{RT_\mathrm{sat}(p)}{Lp} (1-\frac{c_pT_\mathrm{sat}(p)}{L})\omega
\label{eqn:MassContinuitySteamLimit}
\end{equation}
The factor in parentheses multiplying $\omega$ on the right hand side of Eq. \ref{eqn:MassContinuitySteamLimit}
is positive for every plausible atmospheric gas we have checked, so long as one stays away from the critical
point where the formulation becomes invalid.  This implies that $\omega$ has an exponentially decaying
character with increasing altitude (decreasing $p$) and it is not hard to show that, on account of the $1/p$
factor on the right hand side, $\omega \rightarrow 0$ as $p\rightarrow 0$.  Thus, it is possible to satisfy
the $\omega = 0$ boundary condition at the top of the atmosphere even if there is ascent $\omega < 0$
below
\footnote{The equation becomes inconsistent if there is subsidence, because precipitation happens only
on ascent; negative precipitation on subsidence would require the presence of enough retained condensate
to sustain evaporation in the subsiding air mass.}. 
In contrast to the dilute case, in which ascent in a precipitating column must be balanced by subsidence
in surrounding nonprecipitating columns, the mass budget can be satisfied locally in an isolated
ascending precipitating column. As the pure steam limit is approached, the familiar Hadley 
circulation collapses to a single column in which the upward mass flux of vapor is balanced locally by the downward mass
flux carried by precipitation.  Such a "Hadley cell" doesn't transport heat, 
because heat transport in the familiar dilute Hadley circulation
is mediated by compressional heating in the subsiding branch.  As nondiluteness increases
and the pure steam limit is approached, the subsidence and associated heat transport
can be expected to weaken. Further, when the circulation collapses to a single column, it no longer
transports angular momentum horizontally, and therefore does not engage the angular momentum balance,
nor generate subtropical jets, in the way a true Hadley circulation does.  In this degenerate limit,
the dynamics is so different that the circulation can no longer be considered to be a Hadley cell at all. 

An isolated ascending column corresponds to an atmospheric collapse solution.  At
the ground $\omega = dp_s/dt <0$, implying a continual loss of atmospheric mass
associated with precipitation reaching the ground. This is a perfectly valid solution for any value of $\omega(p_s)$,
and additional physical processes must be brought into the picture to remove
the degeneracy and fix a unique equilibrium solution. For example, evaporation
from the surface reservoir would generally adjust until the rate of adding mass at
the bottom of the atmosphere equals the mass loss due to precipitation; in a model,
this would typically be implemented as a thin layer of negative precipitation at the bottom,
within which $\omega$ adjusts from its negative value aloft to zero at the ocean interface.
In a 2D or 3D circulation, mass could also be resupplied by horizontal convergence.   

Integrating Eq. \ref{eqn:MassContinuity} vertically over the whole depth of
the atmosphere yields an equation for the evolution of the surface pressure identical
to the usual one except for the introduction of pressure change due to vertically
integrated $P-E$. When implementing a general circulation model,
this precipitation term, and the corresponding one in Eq. \ref{eqn:MassContinuitySteamLimit},
only needs to be taken explicitly into account for large scale precipitation. For
convective precipitation, the effect of precipitation on $\omega$ is implicitly
taken into account as part of the re-labeling of pressure levels that is generally
necessary in a nondilute convection parameterization, as discussed in \citet{FengNondiluteColumnModel}

When computing the geopotential $gZ$ from the hydrostatic relation, care must
be taken in nondilute atmospheres to account for the effect of the condensible
on atmospheric density. This is a straightforward modification, which is already
incorporated in many conventional general circulation models through use
of a virtual temperature, though in many cases diluteness is assumed when
computing the modification of density due to water vapor.   



\subsubsection{Barotropic nature of strongly nondilute circulations}
\label{section:BarotropicNature}

Section \ref{section:WTG} shows that the atmospheres of slowly rotating planets are generally
expected to exhibit weak horizontal tempreature gradients regardness of their diluteness.
It is abundantly clear that dilute atmospheres on rapid rotators support strong
temperature gradients, as in Earth's extratropics. What is the situation for
nondilute rapid rotators?

\begin{figure}[h]
\centering\includegraphics[width=2.5in]{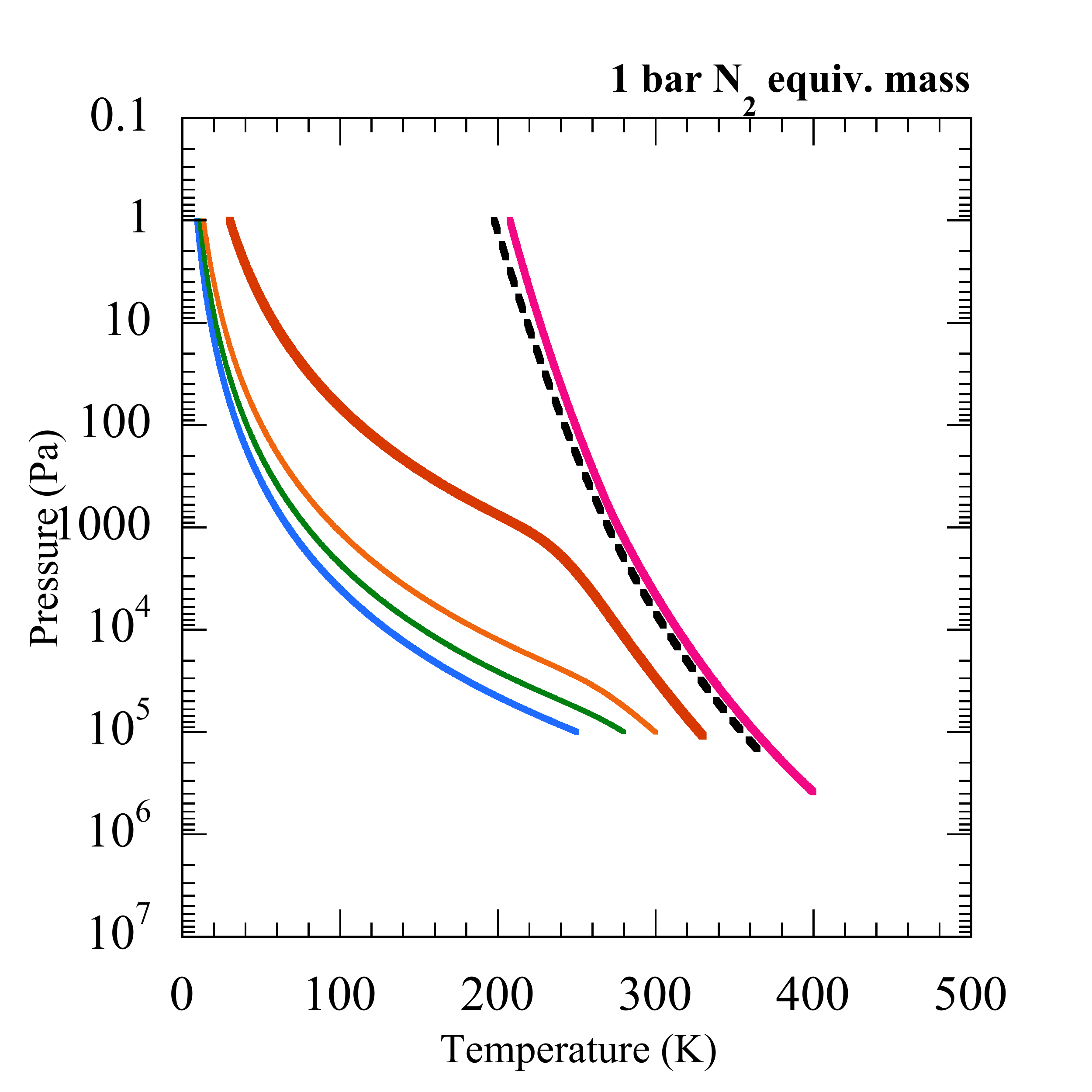}
\centering\includegraphics[width=2.5in]{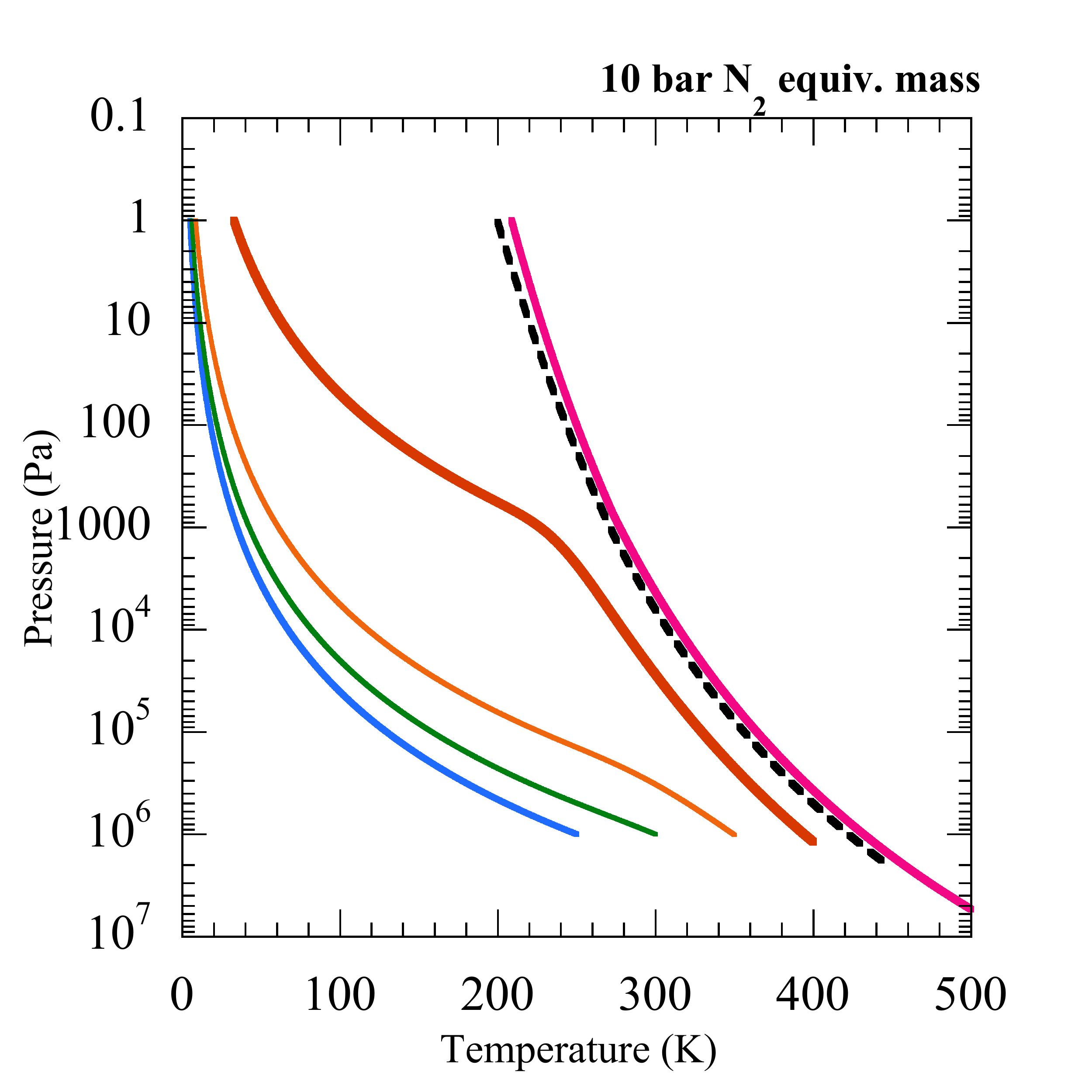}
\caption{A family of saturated moist adiabats with condensing water vapor. The different adiabats
in each panel correspond to different surface temperatures. The left panel shows results for
a 1 bar inventory of noncondensing $\mathrm{N_2}$ and the right shows results for a 10 bar inventory of noncondensing $\mathrm{N_2}$,
In each case, the total atmospheric mass of $\mathrm{N_2}$ is held fixed as the surface temperature is increased.  }
\label{fig:Nondiluteness}
\end{figure}

For sufficiently rapid rotators, the atmospheric flow will be in geostrophic balance.
For geostropically balanced flow, the thermal wind relation links vertical gradients
in wind to horizontal density gradients taken on an isobaric surface. Specifically
introduce a local Cartesian coordinate system $(x,y)$ on a pressure surface, and let
$u$ be the wind in the $x$ direction.  Then the thermal wind relation states
\begin{equation}
\partial_p u = \frac{1}{p f} \partial_y \overline{R}T |_p
\label{eqn:ThermalWind}
\end{equation}
where $\overline{R}$ is the gas constant for the mixture of condensible and noncondensible gas at each point
and $f$ is the local Coriolis parameter.
However, we know that in the saturated
steam atmosphere limit, $T = T(p)$, i.e. temperature is uniform on isobaric 
surfaces within the saturated part of the atmosphere. Moreover, in a pure steam atmosphere
$\overline{R} = R_c = const.$.  Eq. \ref{eqn:ThermalWind}
then implies that the wind is independent of height, and that the dynamics is
barotropic.  Figure \ref{fig:Nondiluteness}, which gives the saturated moist adiabat
for a family of atmospheres with a noncondensible component, shows that as temperature
increases and the atmosphere becomes increasingly nondilute, all the adiabats
collapse onto the pure-steam adiabat.  In consequence, strongly nondilute saturated atmospheres
will tend to have nearly barotropic dynamics. They will have weak horizontal temperature
gradients on pressure surfaces, even if the planet rotates rapidly. 

Note that the above result assumes saturation. Strongly subsaturated atmospheres,
even if pure steam, can be baroclinic.  Hence the factors determining subsaturation
also have a major effect on the dynamics of the atmosphere. We will see some examples
of the behavior of subsaturation in 3D nondilute atmospheres in Section \ref{section:SubsaturationDynamics}. 

The barotropic nature of the system does not in itself preclude the existence of
strong temperature gradients at the planet's surface (if it has one), because the
planetary surface is not an isobaric surface.  The surface pressure gradients are
equal to the pressure gradients aloft, and geostrophically balance the barotropic
winds. Differential stellar heating of the planets surface would lead to surface
temperature gradients, just as the Earth is generally colder at the poles than at the
equator.  On a planet with a surface condensible reservoir, these temperature
gradients would lead to corresponding surface pressure gradients via  the Clausius-Clapeyron
relation.

\begin{figure}[h]
\centering\includegraphics[width=2.5in]{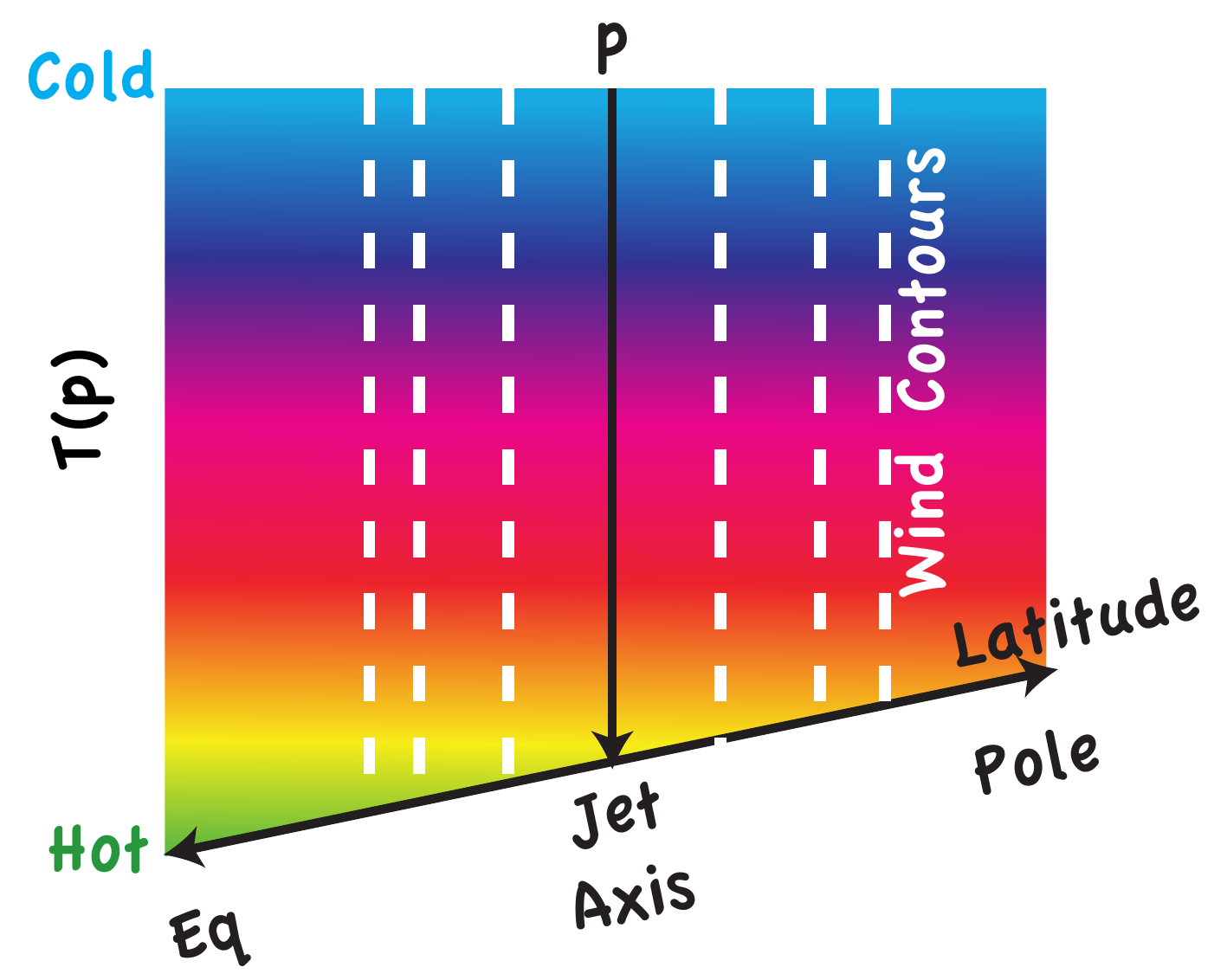}
\caption{latitude-height profile of a barotropic atmosphere with a nonzero surface temperature and pressure gradient.}
\label{fig:BaroWind}
\end{figure}

In a dilute rapidly rotating atmosphere like the Earth's, baroclinic instability
acts to weaken temperature gradients in midlatitudes and the Hadley cell weakens
them in the tropics.  For the saturated strongly nondilute case, the dynamics is barotropic,
so baroclinic eddies cannot play a role in transporting heat. 
Hadley cells are also unable to transport heat efficiently in nondilute atmospheres,
for the reasons discussed in Section \ref{section:Analytic}\ref{section:thermo}\ref{section:MassBalance}. 
However there are a number of other mechanisms that can act to wipe out
surface temperature and pressure gradients in a nondilute atmosphere.

\subsection{Mechanisms of horizontal homogenization of temperature}
\label{section:TempHomog}


\subsubsection{Barotropic adjustment}
\label{section:baro}

Suppose that differential stellar heating of a planet gives rise to a latitude-dependent
surface temperature $T_s(\phi)$. On a planet with a global surface reservoir of condensible,
this will give rise to a surface pressure $p_s(\phi) \approx p_\mathrm{sat}(T_s(\phi))$
if the planet is warm enough for the atmosphere to be dominated by the condensible. 
Because of the exponential dependence of saturation vapor pressure on temperature, a 
small surface temperature gradient translates into a large surface pressure gradient,
and hence (by geostrophic balance) a very large barotropic wind.  Unless the temperature
gradient is sufficiently small, the resulting jet will be barotropically unstable,
and the resulting instability will act to weaken the jet and to redistribute mass
such that the surface pressure and temperature gradients weaken. 

The assumption that barotropic instability acts to adjust the state to one of barotropic
neutrality can be used to obtain an estimate of the surface pressure and temperature
gradient towards which the system is relaxing.  Introduce a local Cartesion coordinate $y$ 
in the vicinity of latitude $\phi_0$, measuring distance in the meridional direction. Let
$u(y)$ be the zonal wind in a geostrophically balanced state.  Then according to the Rayleigh-Kuo criterion neutrality to barotropic instability requires
\begin{equation}
\beta - \frac{d^2u}{dy^2} = 0
\label{eqn:RayleighKuo}
\end{equation}
where $\beta \equiv \frac{df}{dy}(\phi_0) = 2\frac{\Omega}{a} \sin\phi_0$. Geostrophy
and Clausius-Clapeyron imply
\begin{equation}
u = - \frac{1}{f_0\rho}\frac{dp_\mathrm{sat}}{dy} = - \frac{1}{f_0} L\frac{1}{T_s} \frac{dT_s}{dy}
\label{eqn:GeoBalance}
\end{equation}
If we stipulate that winds are weak near the poles,  Eq. \ref{eqn:RayleighKuo} implies $u \sim \beta a^2$, where $a$ is the radius of the planet. Further, the planetary scale temperature variation is $\Delta T_s \sim a\frac{dT_s}{dy}$.  Substituting into Eq. \ref{eqn:GeoBalance} and solving for the temperature variation yields 
\begin{equation}\label{eqn:MoistWTG}
\frac{\Delta T_s}{T_s} \sim \beta f_0 a^3 \frac{1}{L} \sim \frac{\Omega^2 a^2}{L}
\end{equation}
Therefore, the constraints imposed by barotropic stability only guarantee weak temperature gradients for slow rotators; sufficiently rapid rotators can support large temperature
gradients in strongly nondilute atmospheres without the corresponding jet becoming barotropically unstable.  Because the latent heat $L$ is typically a large number,
however, weak gradients can prevail up to quite large values of $\Omega a$.  For example,
for an Earth-sized planet with half Earth's rotation rate, a saturated pure water vapor
atmosphere would have $\Omega^2 a^2/L = .021$, corresponding to roughly a 2\% relative
variation in surface temperature. The expression in Eq. \ref{eqn:MoistWTG} is similar in form to the parameter $\Lambda^{-2}$ which is
small when the WTG approximation is valid, except that $RT$ is replaced by $L$. Since
$L \gg RT$ away from the critical point of the atmospheric gas, the barotropic adjustment process in a saturated 
pure steam atmosphere can maintain weak temperature gradients at higher rotation rates than would be possible in
the absence of condensation.  

\subsubsection{Ekman spin-down}
\label{section:Ekman}

\begin{figure}[h]
\centering\includegraphics[width=3.5in]{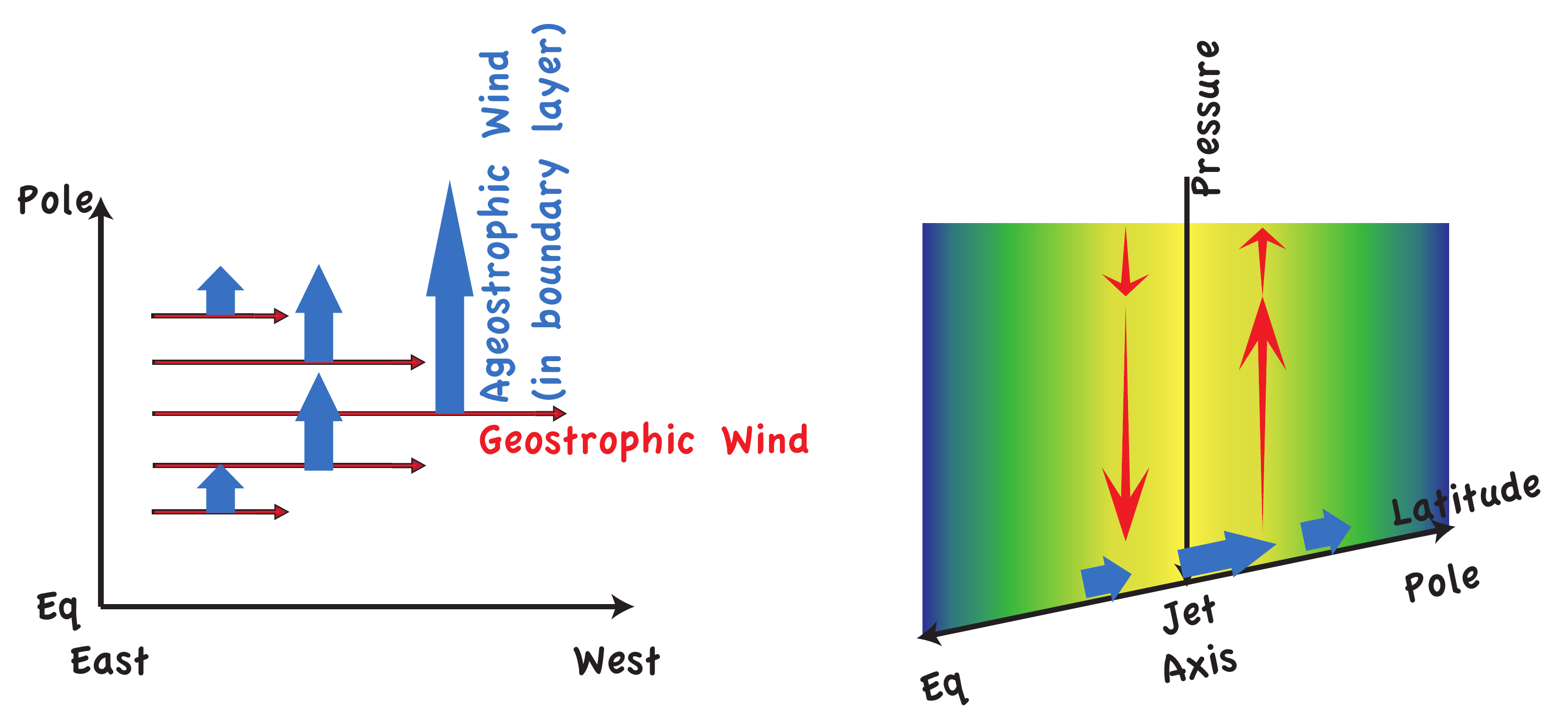}
\caption{Schematic of transport due to Ekman drift in a frictional boundary layer. The left panel shows
the geostrophically balanced horizontal wind (thin arrows) and corresponding frictionally driven
ageostrophic boundary layer wind (thick blue arrows) in the longitude-latitude plane.  The right panel shows
the cross section in the latitude-pressure plane (high pressure toward the bottom) of the geostrophic barotropic 
jet (speed indicated by color shading), together with the ageostrophic boundary layer wind (blue arrows) and
vertical wind arising from convergence/divergence of the ageostrophic wind (red arrows).}
\label{fig:Ekman}
\end{figure}

For planets with a surface which exerts frictional drag on the overlying atmosphere, Ekman spin-down offers another way 
of damping surface pressure and temperature gradients.  Surface temperature gradients induce surface pressure gradients via
Clausius-Clapeyron, and these gradients are in turn balanced by a geostrophic wind.  Within the frictional boundary layer,
the dominant balance between frictional forces on the wind and the Coriolis force induces a drift in the boundary layer
in the direction from high pressure to low pressure, as illustrated in Fig. \ref{fig:Ekman}. Alternately, the process
can be understood in terms of the "Einstein's Teacup" analogy, in which the pressure gradients in the boundary layer 
are nearly the same as in the geostrophically balanced atmosphere aloft, but friction has weakened the wind that induces
the balancing Coriolis forces, thus leaving an unbalance pressure gradient that drives a downgradient current. With either
way of looking at it, the process spins down the wind, transport heat and mass toward cold regions, and acts to damp the surface pressure and temperature gradients. 

For an Earth-like fast-rotating planet, the Ekman layer is thin enough that the atmospheric temperature in the layer is close to the surface temperature ($T_s$). Then the energy balance of the planetary system (ignoring heat flux in the ocean) at latitude $\phi$ can be simplified as
\begin{equation} \label{eq:energy_pure}
S_{abs} (\phi) - OLR(T_s(\phi)) \approx \frac{1}{a \cos \phi} \frac{\upd}{\upd \phi} \Big( \cos (\phi) M_E(\phi) L(T_s) \Big)
\end{equation} 
where $S_{abs}$ is the absorbed stellar flux and $OLR$ is the outgoing longwave radiation of the planetary system respectively, 
$a$ is the radius of the planet, $M_E$ is the meridional Ekman mass transport and $L$ is the specific latent heat of vaporization (assumed to
have a weak temperature dependence). The left hand side of Eq.(\ref{eq:energy_pure}) represents the net downwelling radiative flux at the top of the atmosphere and the right hand side represents the divergence of the meridional energy flux. Assuming that friction in the Ekman layer is parameterized by a constant eddy viscosity $A$, $M_E$ can be described by the product of the layer thickness  and the geostrophic zonal wind (\citet{vallis}, p. 112), and then is related to the meridional gradient of surface temperature through the Clausius-Clapeyron relation,   
\begin{equation} \label{eq:ekman}
M_E \approx \frac{1}{2} \sqrt{\frac{A}{\Omega \sin \phi}} \frac{1}{2 \Omega a \sin \phi} \left( -  \frac{\upd p_{sat}(T_s)}{\upd \phi} \right)
  =\frac{1}{2} \sqrt{\frac{A}{\Omega \sin \phi}} \frac{1}{2 \Omega a \sin \phi}\left(\frac{Lp_{sat}(T_s)}{RT_s^2}\right) \left( -  \frac{\upd T_s}{\upd \phi} \right)
\end{equation}
where $\Omega$ is the spin rate of the planet. The final equality proceeds from application of the chain rule and Clausius-Clapeyron.  Furthermore, the outgoing longwave radiation (OLR) of a steam atmosphere is nearly a linear function of $T_s$ before the runaway greenhouse occurs. With the above simplifications, Eq.(\ref{eq:energy_pure}) becomes the equation of a diffusive energy balance model with a latitude-dependent diffusivity. The steady-state meridional distribution of $T_s$ can be obtained by solving such a diffusion equation numerically given the insolation profile. 

We can simplify Eq.(\ref{eq:energy_pure}) further by the mass balance equation if considering the limit in which the horizontal energy transport is so efficient that the global surface becomes isothermal (denoted as $T_{s0}$),
\begin{equation} \label{eq:mass_pure}
S_{abs} (\phi) - OLR(T_{s0}) \approx \frac{1}{a \cos \phi} \frac{\upd}{\upd \phi} \Big( \cos (\phi) M_E(\phi) \Big) L(T_{s0}) = (E-P) \Big|_\phi L(T_{s0})
\end{equation} 
where $(E-P)$ is the net evaporation rate at the surface ($E$ represents the evaporation and $P$ the precipitation) and is proportional to the net radiative flux at the top of the atmosphere. This implies the condensable substance goes from the surface reservoir to the atmosphere in the low-latitudes, and returns the reservoir again in the high-latitudes. 

In Section \ref{section:TempAndHumiditySims} we will show simulations demonstrating the effectiveness of this mechanism for equalizing 
surface pressures and temperatures in nondilute atmospheres.

\subsection{Inhibition of subsaturation for strongly nondilute flow}
\label{section:SubsatInhibition}

Much of the discussion of the novel behavior of nondilute atmospheres relies on
the assumption that nearly saturated conditions prevail within extensive regions
of the atmosphere.  In Earth's present dilute atmosphere, in contrast, most of
the atmosphere is highly undersaturated, even in the free troposphere over the
oceans.  In Earth's dilute atmospehre, substuration is produced primarily through
transporting moist air parcels to colder, lower-pressure places (poleward or aloft)
where adiabatic expansion and other cooling mechanisms cause most of the condensible
substance to rain out, and then transporting them back to higher-pressure warmer places
where the warming increases the saturation vapor pressure of the parcel but not the
condensible content (until it picks up a new supply of condensible from some source)
\citep{pierrehumbert07}.  This can happen at the scale of baroclinic synoptic eddies
in midlatitudes, or on global scales due to planetary Rossby waves in the midlatitudes
or Hadley and Walker circulations in the tropics. It can also happen at the scale
of ensembles of convective clouds. 

There are a number of factors that make it much harder for strongly nondilute atmospheres
to generate significant subsaturation.  Once a saturated layer is established in such
an atmosphere, the dynamics becomes largely barotropic, as discussed in Section \ref{section:BarotropicNature}, which precludes the vertical motions needed to 
generate subsaturation.  Thus, saturated layers tend to maintain themselves in strongly
nondilute atmospheres. Further, the mass-flux in the upward branches of Hadley and Walker
circulations do not need to be fully compensated by subsidence in the surroundings in
strongly nondilute atmospheres, as the upward flux of vapor is mostly balanced {\it in situ}
by the downward mass flux associated with precipitation (Section \ref{section:MassBalance}).
In dilute atmospheres with extensive tropical circulations, the large scale subsidence
which brings down dry air from the upper troposphere is a key source of subsaturated air
\citep{pierrehumbert07}.

More generally, subsaturation is difficult to produce in strongly nondilute atmospheres
because a significant isobaric gradient in subsaturation between neighboring air parcels requires
a gradient in temperature, since in the pure steam limit the degree of saturation corresponds to
the excess of temperature over $T_\mathrm{sat}(p)$. In contrast, for a dilute atmosphere, a saturated air parcel
can have the same temperature (more precisely the same virtual temperature) 
as a neighboring completely dry parcel.  In a nondilute
atmosphere the temperature gradients associated with a subsaturation contrast extending
over a deep layer would lead to strong pressure gradients in accordance with hydrostatic
balance.  This would drive circulations that tend to mix away the gradient.  
By itself this only implies weak gradients
in degree of subsaturation, but if there is any portion of the atmosphere that is maintained
near saturation (e.g. in a region of strong nondilute convection), the saturation
there will tend to be transmitted to the rest of the atmosphere. 

The preceding effects inhibit dynamic production of subsaturation of the type that
occurs in Earth's troposphere, but subsaturated layers can nonetheless be produced
by radiative means in a nondilute atmosphere.  Any layer of the atmosphere within
which the local radiative equilibrium $T_\mathrm{r.e}(p)$ exceeds the saturation
temperature $T_\mathrm{sat}(p)$ will be subsaturated.  Such layers would typically
be produced by strong absorption of shortwave stellar radiation by the atmosphere,
and if the atmosphere is so strongly absorbing that essentially no shortwave
radiation reaches the surface, the subsaturated layer can even extend all the way
to the condensed reservoir at the surface. Note that the stellar heating need not
be strong enough to create a temperature inversion. Temperature can decrease
with height, so long as it doesn't fall below $T_\mathrm{sat}(p)$.  If the
atmosphere is very optically thick in the infrared, however, even a trickle
of shortwave radiation reaching the surface would heat the surface enough
to be warmer than the overlying radiative-equilibrium temperature, leading ultimately
to a saturated precipitating layer extending upward from the surface. 

Although the above arguments lend a certain plausibility to the assertion
that atmospheres should become more saturated as they become more nondilute
(barring effects of strong internal radiative heating), the dynamical issues
are subtle and difficult to quantify convincingly without some guidance from
full 3D simulations.  We'll return to the issue of subsaturation in Section \ref{section:SubsaturationDynamics}.

\section{Description of the Exo-FMS model}
\label{section:FMSdescription}

We have developed an idealized three-dimensional general circulation model (3D GCM) with simplified physical parameterizations including an active hydrological cycle based on formulations that remain valid regardless of the diluteness of the condensable substance in the atmosphere. It is a variant of the model used to study idealized atmospheres  in \cite{ogorman} and \cite{frierson} and the model used to study tidally-locked Earth-like aquaplanets  in \cite{merlis}. These  models are designed to improve our understanding of fundamental processes in complex planetary atmosphere, and are different from the models of a high degree of complexity used for making precise predictions. 

Two major modifications are made in our model compared with previous studies. First, to avoid numerical instabilities at high concentrations of the condensable substance, our model (hereafter referred to as Exo-FMS) uses the finite-volume dynamical core of the GFDL Flexible Modeling System (FMS) described in \cite{lin} instead of the spectral transform method used in  \cite{ogorman,merlis} 
to solve three-dimensional primitive equations of an ideal-gas atmosphere. Second, Exo-FMS uses an energy-conserving simplified moist convection scheme without any approximation  about the diluteness of the condensable substance while conventional GCMs usually use the dilute approximation and assume the surface pressure and thermodynamic parameters of the atmosphere is insensitive to the amount of condensible substance in the atmosphere. 
Our moist convection scheme is similar in spirit to the hard convective adjustment scheme in \cite{manabe}, but
conserves non-dilute moist enthalpy while the condensate is produced and retained in the atmosphere, and takes the energy change due to mass redistribution into account when the condensate falls to the surface or enters the atmosphere through surface evaporation. The condensate falls within the same time step right after it is produced in the super-saturated layer. The nondilute moist convection scheme is described and validated in \cite{FengNondiluteColumnModel}, where we show that it conserves both energy and mass from the dilute to non-dilute regime. Similarly, our large-scale condensation scheme takes into account the conservation of non-dilute moist enthalpy and the energy loss associated with large-scale condensation precipitation.   Microphysical effects are no doubt of considerable importance in all planetary atmospheres with
a condensible component, but in these simulations we neglect the microphysics governing formation and removal of
precipitation, as well as effects of retained condensate and atmospheric heating due to frictional dissipation surrounding
falling precipitation. 

Since we wish to highlight dynamical phenomenon in the present paper, the Exo-FMS simulations presented here use a gray radiation scheme in the infrared spectral region, and assumes the atmosphere is transparent to shortwave radiation. The longwave optical depth $\tau$ has two contributions
\begin{equation}
\tau(p) = \kappa_0 \int_{0}^{p} q(p') \frac{\upd p'}{g} + \tau_1 \frac{p}{p_s}
\end{equation}
where $\kappa_0$ and 
$q$ are the absorption cross-section and the mass concentration of the condensable substance respectively, and $\tau_1$ is the total optical thickness of the non-condensable absorber which is assumed to be well-mixed in the atmosphere . Both $\kappa_0$ and $\tau_1$ are constants.
In the simulations we present in this paper, $\tau_1$ and $\kappa_0$ are independant of the total mass of noncondensable in the atmosphere.
This is somewhat unrealistic since it assumes the total mass of the noncondensable greenhouse gas decreases with $p_s$ in such a way as to
compensate for pressure broadening and keep $\tau_1$ fixed, and also neglects pressure broadening or collisional opacity of the 
condensable component.  However, this experimental protocol allows us to keep the mean surface temperature roughly fixed as 
the mass of the noncondensable background is changed, and thus makes it easier to isolate the effects of nondiluteness. 
For similar reasons we avoid the complex issues associated with cloud feedbacks by neglecting the radiative effect of clouds. 
 
The opposite regime from the shortwave-transparent atmospheres we consider in the present work is the one in which atmospheric
shortwave absorption is so strong that essentially all incoming stellar radiation is deposited directly in the atmosphere, rather
than being deposited at the ground and being communicated upwards by radiation and convection.  Many exoplanets, including low density
Super-Earths such as GJ1214b, are expected to be fluid planets with no distinct surface at all. Planets without appreciable deposition
of stellar energy at a distinct surface can nonetheless have nondilute layers, as discussed in Section \ref{section:WhenNondilute}.  This is an important regime, which will be the subject of future work.

In Exo-FMS, the specific heat capacity of the substances in the atmosphere are treated as constants, but the specific latent heat of vaporization has a weak temperature dependence. 
The specific latent heat of a material varies approximately linearly with the temperature when the density of the condensate is much greater than that of its vapor phase  (\cite{emanuel}, p. 115).
\begin{equation}
\frac{\upd L}{\upd T} \approx c_{pc} - c_{p\ell}
\end{equation}
Here $c_p$ is the specific heat capacity at constant pressure, and
subscript ``$c$'' and ``$\ell$'' represents the vapor phase and the condensate, respectively. With this form of temperature dependence
it is fairly straight forward to formulate a moist enthalpy incorporating both gaseous and condensed phases, which is conserved when
the enthalpy carried by precipitation is taken into account.  For simplicity, there is only a single phase transition in Exo-FMS,
which could equally be interpreted as liquid-vapor or ice-vapor. 

The lower boundary of the column model is a liquid slab ocean with uniform thickness. At the surface, the sensible and latent heat fluxes are computed based on the bulk exchange formulae  (see \cite{ClimateBook}, p.~396) assuming constant drag coefficient ($C_D = 0.005$). The surface pressure is calculated not only by the air flow divergence above the ground as in  conventional GCMs, but also by the $(E-P)$ flux at the surface. In the surface energy budget, other than the sensible and latent heat fluxes and the radiative flux, we also take into account the internal and potential energy flux of the exchanged mass between the atmosphere and the surface reservoir to conserve energy. 
In  Exo-FMS, both the  mass  of the non-condensable and condensable substances (including both phases in the atmosphere and surface reservoir) are conserved. The former conservation is  automatically achieved by the FMS finite-volume dynamical core. The mass budget of the slab ocean requires a mass redistribution scheme to even out the horizontal variations of the ocean depth created by the distribution of $P-E$. Otherwise, the liquid substance will accumulate in some places and be depleted in other places. At each time step, mass is redistributed so as to keep the ocean depth globally uniform, and then a
globally uniform temperature offset is applied to the ocean temperature so as to enforce conservation of oceanic energy
in the course of the mass adjustment. The oceanic heat transport implied by the temperature adjustment is small because it
is only the heat associated with the mass added to or taken away from an ocean column that is mixed throughought the global
ocean, leaving the pre-existing temperature gradients intact. 
Details of the adjustment scheme, as well as other technical details concerning the formulation of Exo-FMS, are provided
in the Supplementary Materials. 

\begin{figure}[!h]
\centering
\includegraphics[width=\textwidth]{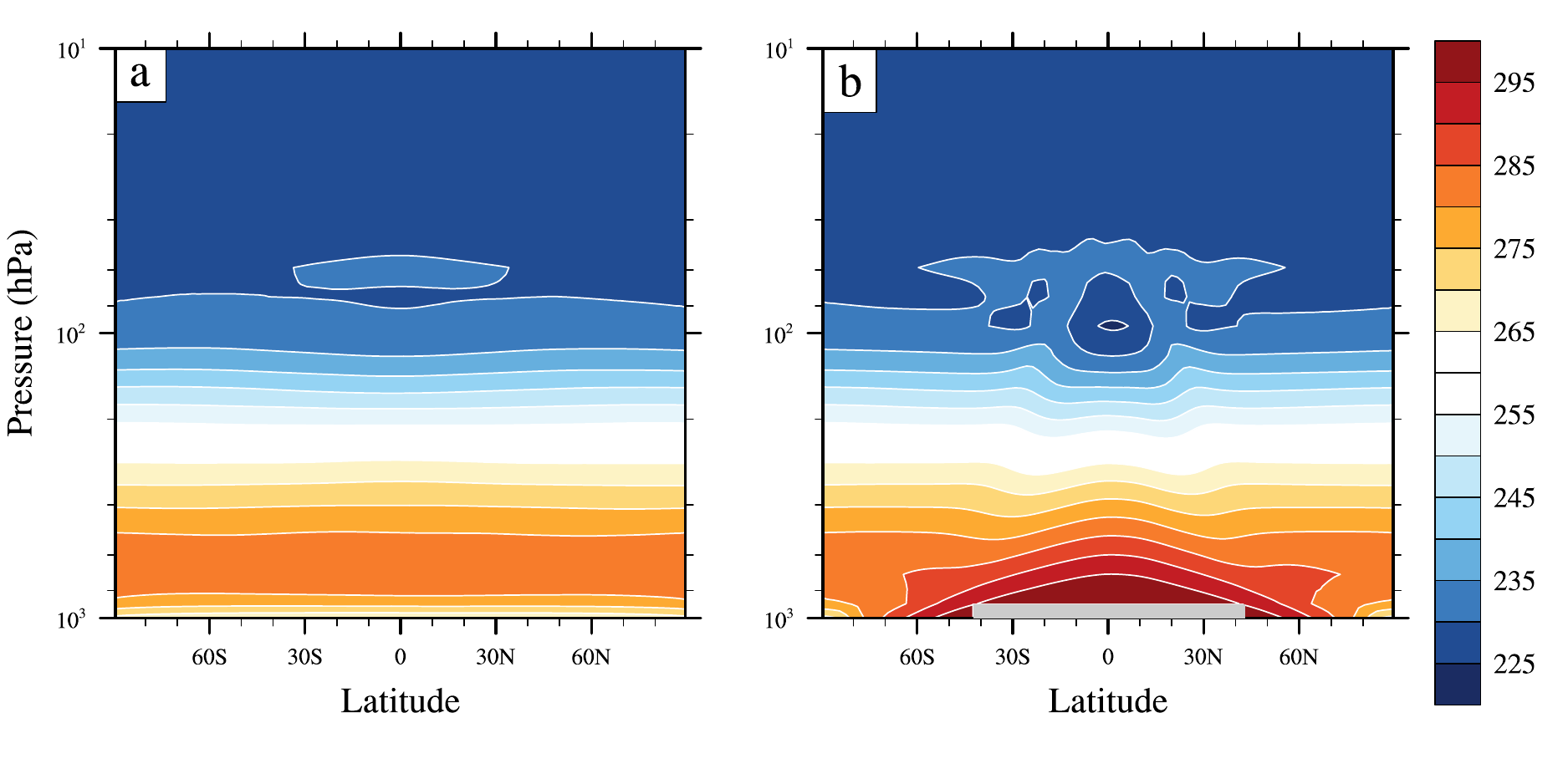}
\caption{Cross section of the atmospheric temperature (K)  along anti-stellar longitudes (a) and substellar longitudes (b) of a tidally-locked simulation conducted in the dilute regime.  Averages are taken over 10$^\circ$ of longitude.}
\label{fig1_tidal_t}
\end{figure}

\begin{figure}[!h]
\centering
\includegraphics[width=\textwidth]{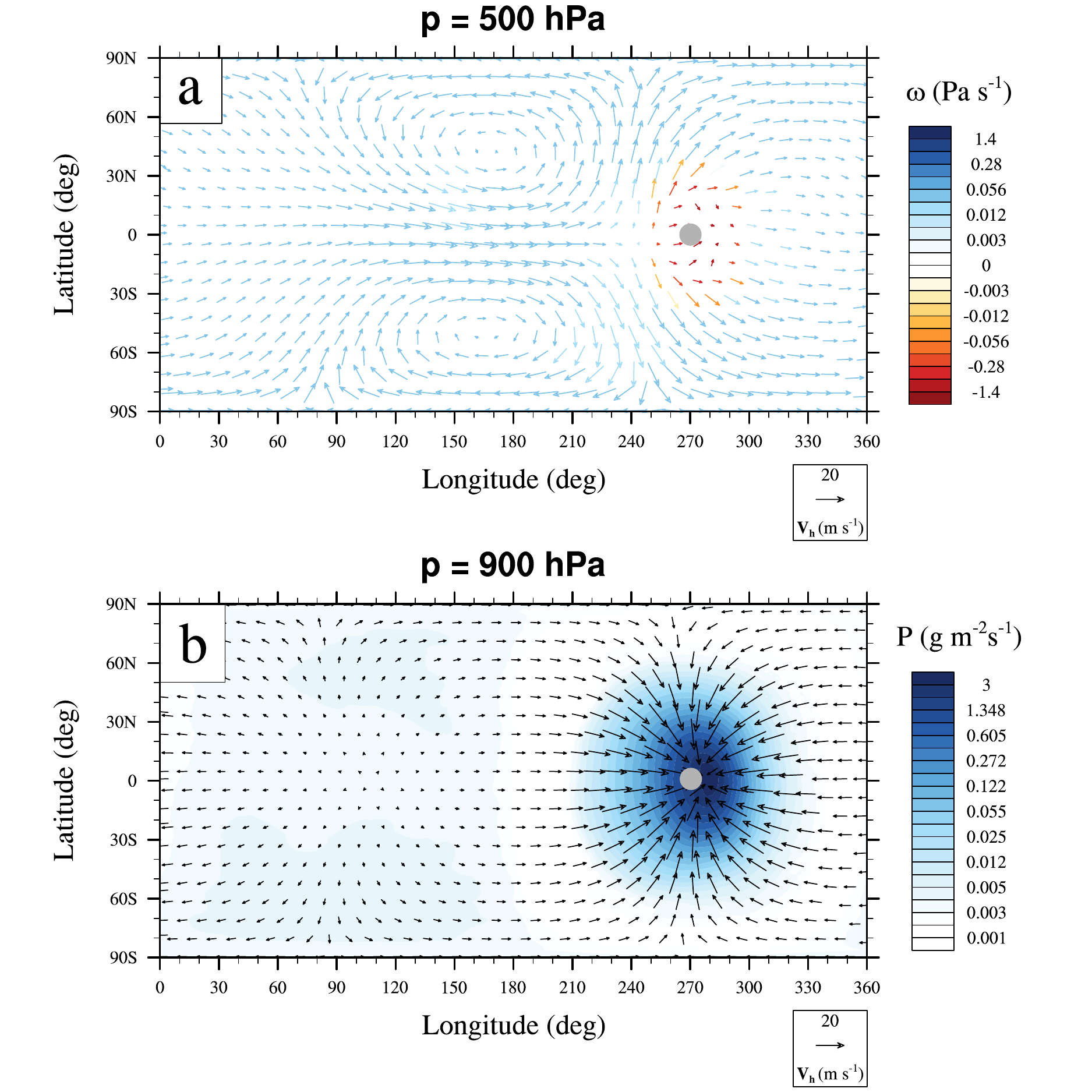}
\caption{Upper panel: 500 hPa horizontal winds (arrows) and vertical pressure velocity $\omega$ (arrow colors) for the dilute tidally locked simulation. Lower panel: 900 hPa winds (arrows) and net column precipitation (color shading), The grey circle marks the substellar point. }
\label{fig1_tidal_wind}
\end{figure}

To show that Exo-FMS is capabable of reproducing features familiar from other moist simulations of circulation of tide-locked planets, we will show one Earth-like tidally locked simulation with experimental design essentially identical to those described by \cite{merlis}, except that the period of the circular orbit is taken to be 50 Earth days. Two latitude-height cross sections of temperature are shown in Fig. ~\ref{fig1_tidal_t}.  Exo-FMS reproduces the WTG behavior (Figure~9 in \cite{merlis}) in the free troposphere.  In our simulation, the horizontal temperature gradient is small aloft both on the dayside and nightside of the planet. Other familiar features include the shallow hot-spot in the lower atmosphere
over the substellar point, and the nightside temperature inversion near the surface.  Horizontal winds, vertical velocity and precipitation are shown in Fig. ~\ref{fig1_tidal_wind}, and reproduce the typical features of strong low-level convergence into a region of concentrated updraft and precipitation near the substellar point, subsidence aloft over most of the planet.  Aloft, the simulation reproduces the familiar 
equatorial super-rotating winds and the global Kelvin/Rossby wave pattern expected in a planet with weak but nonvanishing rotation. 

\section{3D simulations of nondilute atmospheres on rapid rotators}
\label{section:FMSRapidRotSims}

\subsection{Simulation design}
\label{section:SimDesign}

Since the novel effects of nondiluteness are most striking in the case of rapid rotators,
we concentrate on that case here.  Hot tide-locked planets in short-period orbits can be
rapid rotators, but we'll also focus here on Earthlike annual mean instellation patterns, which
are symmetric about the rotation axis.    For tide-locked planets, whether slowly or rapidly 
rotating in the sense of the WTG approximation, nondiluteness is expected to cause interesting
and important deviations from the dilute pattern consisting of subsaturated subsiding air over
most of the planet, but that will be the subject of a future paper.  

Our simulation design is similar to the study of the runaway greenhouse state of a 3D gray atmosphere in \cite{ishiwatari02}, except
have performed simulations with a range of values of the inventory $p_{a0}$ of the noncondensible gas. The noncondensible inventory
is the parameter we vary in order to control the diluteness of the atmosphere. An alternate way to control diluteness is to keep
$p_{a0}$ fixed but vary the instellation of the planet (corresponding to changing the luminosity of the star or the radius of the orbit),
but varying $p_{a0}$ makes for a cleaner exploration of nondiluteness effects because the maximum surface temperature remains approximately
fixed in our simulation. All other parameters are chosen to mimic the present Earth conditions. 
The non-condensable substance in the simulation is mixed N\2-O\2 air, as in the present Earth's atmosphere, while the condensable substance is water vapor.  As in \cite{ishiwatari02}, the longwave absorption cross-section of water vapor $\kappa_0$ is a constant \footnote{An average angle should be taken into account in the two-stream radiation equations because the infrared flux tends to become forward-peaked as it propagates. Our chosen $\kappa_0$ has already include this effect. This is slightly different from the radiation scheme utilized in \cite{ishiwatari02} who do not specify the average angle they chose.}, 0.01\,$\mathrm{m^2\,kg^{-1}}$. The atmosphere is transparent to the shortwave incoming radiation and the surface albedo is zero.  

The top-of-atmosphere instellation is given by the annual and zonal average evaluated with the present Earth's orbit, approximated by
\begin{equation}
S_{TOA}(\phi) = S_0 (1 + 0.3(1-3\sin^2 \phi))
\end{equation}
where $S_0 = 300\,\mathrm{W\,m^{-2}}$ is the global mean insolation and there is no diurnal or seasonal cycle. 
The horizontal resolution utilized in this study is 144$\times$96, and there are 40 unevenly spaced  vertical levels in hybrid $\sigma = p/p_s$ coordinate (pressure $p$ and surface pressure $p_s$). The pressure of the radiative level in a pure water-vapor atmosphere in our model is 9.8\,hPa. Given that the strongest radiative cooling and convective heating of water vapor occurs near the radiating level, the pressure at the top of our model is chosen to be ~0.01\,hPa in order to resolve the radiative and convective processes in the non-dilute atmosphere.

As listed in Table~\ref{table_pdry}, four experiments are presented with different values of noncondensible inventory $p_{a0}$. 
Each experiment is referred to as a letter P followed by the value of $p_{a0}$ in hPa specified in the experiment. Experiment P1000 is the run that mimics the present Earth's atmosphere. The results we present are averages over the last 600\,days of 3000-day integrations.

\begin{table}[!h]
\caption{The list of experiments and the values of the initial surface partial pressure of the N\2-O\2 air ($p_{a0}$), global mean surface temperature ($T_s$) and global mean mass concentration of water vapor in the atmosphere $q$ for each experiment.}
\label{table_pdry}
\begin{tabular}{llll}
\hline
Experiment & $p_\mathrm{dry}$  (hPa) & Global mean $T_s$ (K) & Global mean $q$ (kg/kg)\\
\hline
P1000 & 1000 & 294.40 & $6.41\times10^{-3}$ \\
P30 & 30 & 287.37 & 0.21 \\
P5 & 5  &  292.71 & 0.70 \\
P0 & 0  &  287.79 & 1.00 \\
\hline
\end{tabular}
\end{table}

Because the mass of dry air in our experiments has no direct radiative effect, the global mean $T_s$ varies only by 7\,K\footnote{There is an indirect radiative effect through the effect of the dry air on the lapse rate, and through nonlinear rectification of surface temperature variations in latitude.}  (Table~\ref{table_pdry}). Hence, it is primarily the mass of dry air that alters the concentrations of water vapor in the atmosphere by two orders of magnitude. Though we vary the partial pressure of the non-condensible gas in theses simulations primarily as a convenient
means of controlling nondiluteness, there are in reality many real physical processes that can enter into the determination of
the noncondensible inventory on actual planets, so exoplanet climate studies should always consider a wide range of possible values
for this parameter.  
For our own planet Earth, a surface pressure of two or three times the present value is proposed to help reduce the amount of
$\mathrm{CO_2}$ needed to resolve the `Faint Young Sun' paradox,  via pressure-broadened increase of infrared absorption of greenhouse gases \citep{goldblatt09} or the H\2-N\2 collision-induced absorption \citep{wordsworth13}, though fossil 
raindrop imprints suggest an absolute upper limit of less than 2100\,hPa for the late-Archean climate \citep{som12}. The mass of the non-condensable on Earth-like exoplanets is even more difficult to constrain by transit spectra.   The mass of the non-condensable can affect the planetary climate through various ways,  and our study here only focuses only on diluteness effects.

\subsection{Basic description of zonal-mean temperature and humidity fields}
\label{section:TempAndHumiditySims}

\begin{figure}[!h]
\centering
\includegraphics[width=\textwidth]{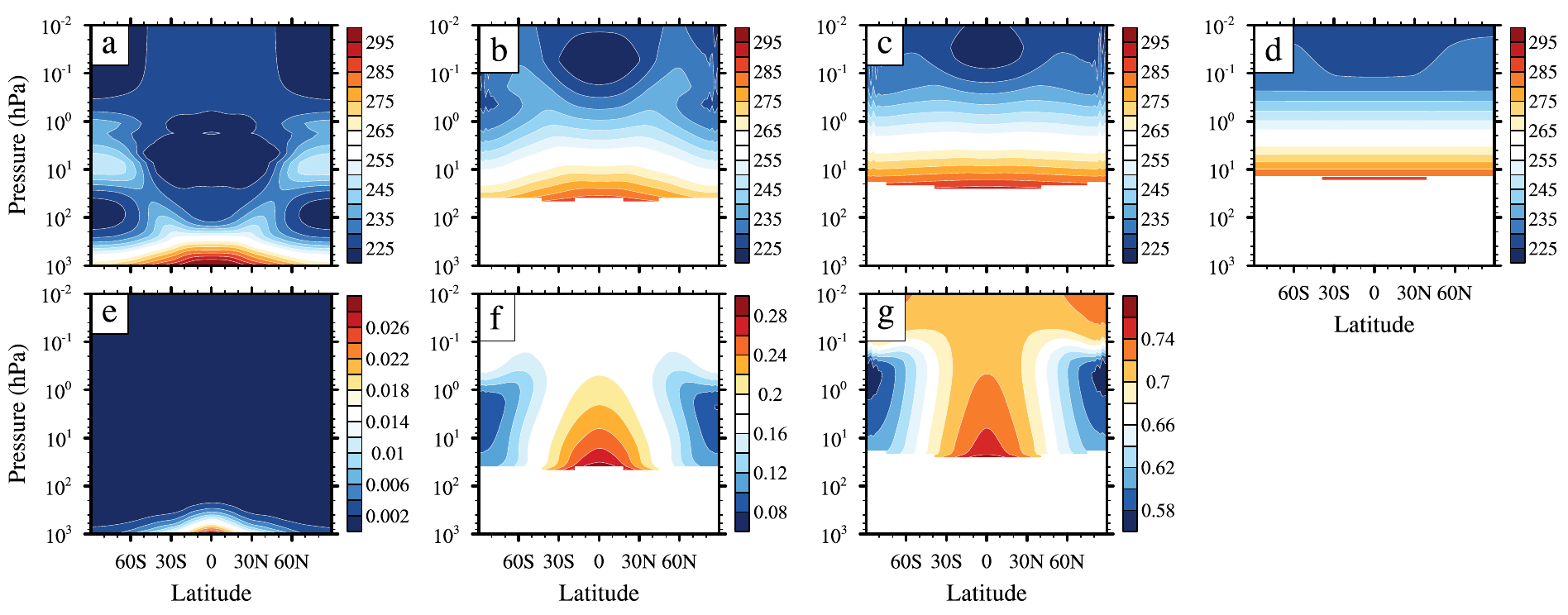}
\caption{Zonal-mean cross sections of the temperature (K) in experiments (a) P1000, (b) P30, (c) P5 and (d) P0, and the mass concentration of water vapor (kg/kg) in experiments (e) P1000, (f) P30 and (g) P5. The concentration of water vapor in experiment P0 is unity everywhere in the atmosphere, and therefore is not shown in the figure. The plots of temperatures share the same contour intervals. The white horizontal bars at the bottom of the atmosphere are due to meridional surface temperature variations when the atmosphere becomes non-dilute.}
\label{fig2_tq}
\end{figure}

The zonal mean temperature fields in the simulations are shown in Figs.~\ref{fig2_tq}a-d, and the specific humidity fields (Figs.~\ref{fig2_tq}e-g)
provide an indication of the degree of diluteness of the atmosphere in each simulation. The most dilute experiment, P1000, exhibits a latitude-height pattern of temperature and humidity similar to the Earth's present atmosphere, with WTG behavior seen only in the tropics.   As the atmosphere is made more nondilute, the vertical temperature gradients become weaker; this is a simple consequence of the flatness of the moist adiabat, and its tendency to become nearly 
isothermal as the steam atmosphere limit is approached, as discussed in Section \ref{section:SteamLimit}. The temperature gradient
on isobaric surfaces also becomes weaker, as expected from the discussion in Section \ref{section:BarotropicNature}. Moreover, the planet's surface becomes nearly isobaric (and hence nearly isothermal), as argued in Section \ref{section:TempHomog}.   
Specifically, the equator-to-pole surface temperature differences in the four experiments reduce from 40\,K in P1000 to 1\,K in P0.

Note that even in the very nondilute case P5, the low level specific humidity varies by nearly 25\%, with lowest values appearing in the polar regions.  This may be surprising in view of the weak temperature gradients, but it is a simple consequence of the fact that the exponential dependence of vapor pressure on temperature makes the specific humidity very sensitive to temperature in cases with just a small amount of
noncondensible gas. 

\subsection{The nature of the circulation}
\label{section:circulation}

\begin{figure}[!h]
\centering
\includegraphics[width=\textwidth]{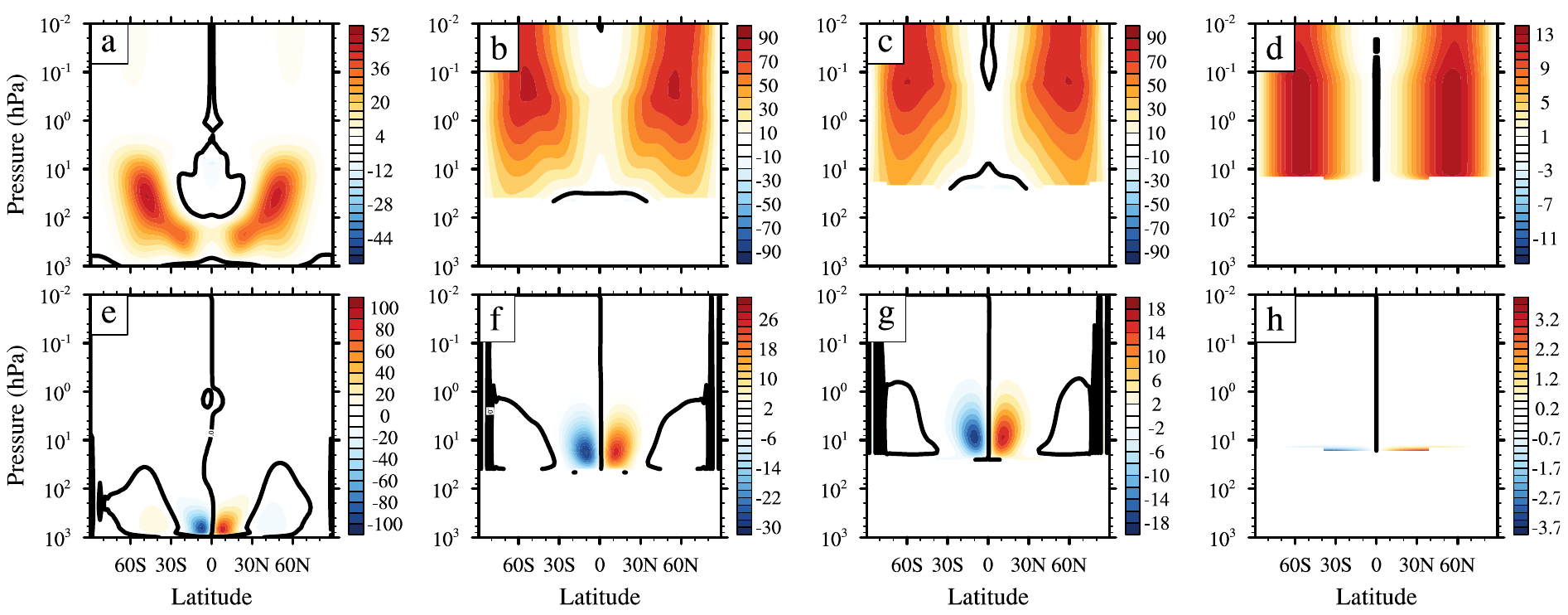}
\caption{Upper panels: Zonal-mean cross sections of the zonal wind component (m\,s$^{-1}$) in experiments (a) P1000, (b) P30, (c) P5 and (d) P0. Low panels: Same as the upper panels, but for the zonal-mean cross sections of the mass streamfunction in 10$^9$\,kg\,s$^{-1}$ in each experiments.}
\label{fig3_upsi}
\end{figure}

Latitude-height profiles of zonal mean wind are shown in Figs.~\ref{fig3_upsi}a-d, with corresponding meridional circulation shown in
Figs.~\ref{fig3_upsi}e-h. The streamfunctions are computed using only the large scale velocity, and not the convective mass transport
of condensible implicit in the convection parameterization, which is balanced by the mass flux of precipitation.   For experiment P1000, which has Earthlike diluteness, there are baroclinic jets associated with midlatitude temperature gradients. Detailed examination of the
transient eddy properties is beyond the scope of the present work, but an examination of the time behavior (not shown) reveals
Earthlike midlatitude baroclinic eddies and eastward-propagating convectively coupled waves in the tropics. There is a strong Hadley
circulation. This only appears shallow because of the large range of pressures plotted; it in fact reaches to approximately 200 hPa.  
Associated with this circulation is a pair of strong subtropical jets.  These subtropical jets are somewhat more strongly separated
from the midlatitude jets than is generally the case in Earth's present climate, but that may be due to the aquaplanet configuration. 

At the other extreme, in the pure steam simulation there is essentially no horizontal temperature gradient in the interior of the atmosphere. In accordance with Section \ref{section:Analytic}\ref{section:thermo}\ref{section:BarotropicNature}, the vertical zonal wind shear vanishes and two barotropic jets emerge at $\sim 60^\circ$ (experiment P0, Figure~\ref{fig3_upsi}d). The jets are weak and barotropically stable, and in fact the system settles into a time-independant
state with no transient eddies at all.  This suggests that the surface pressure gradients are being determined by the Ekman transport 
mechanism rather than barotropic adjustment, which would proceed via barotropic instability.  The Hadley cell has completely vanished,
and meridional mass transport is dominated by boundary layer currents which carry mass from the high pressure tropics to the low pressure poles.

The pure steam simulation illustrates a striking effect of nondiluteness on the angular momentum budget of the planet.  In this simulation,
the low level flow is everywhere poleward, and maintains surface winds that are westerly everywhere.  At first glance, this appears
paradoxical, because the resulting unbalanced torque would spin up the planet without bound, violating angular momentum conservation.
The resolution relates to the angular momentum carried to the surface by precipitation.  The everywhere-poleward mass flux, which would
be impossible in a dilute atmosphere, is possible in this case because the atmosphere picks up mass through evaporation in the tropics
and loses it by precipitation at higher latitudes. The coriolis force acting on this current maintains the global surface westerlies
against friction. The mass added to the atmosphere in the tropics takes high angular momentum with it. As air parcels move poleward,
their angular momentum reduces through the action of friction. Thus, when mass is returned to the surface by precipitation, it has lower
angular momentum than when it left the surface in the tropics.  This acts as a sink of angular momentum of the surface ocean, compensating
the net torque exerted on it. In experiment P0, the global mean value of the angular momentum source and sink are estimated as 5.49$\times10^4$\,N$\cdot$m\,/m$^2$ and 5.60$\times10^4$\,N$\cdot$m\,/m$^2$, respectively, which 
very nearly closes the angular momentum budget.  

In the intermediate cases P30 and P5, the main surprise is that the vertical wind shear is quite strong, despite the weakening horizontal
temperature gradients. In fact, the maximum jet speed actually increases between the Earthlike case and the more nondilute case P30. 
This behavior arises in part because the thermal wind relation (Eq. \ref{eqn:ThermalWind}) has a pre-factor $p^{-1}$ which tends to amplify
the vertical shear (as seen in pressure coordinates) when the pressure is low. The rest of the behavior is due to a novel dynamical effect
of moisture in the nondilute regime.  In a dilute atmosphere, the gas constant $R$ in Eq. \ref{eqn:ThermalWind} is nearly constant. It is
also constant in the pure steam limit. However, in nondilute atmospheres that aren't close to the pure steam limit, $R$ varies greatly
in the horizontal because the variations of specific humidity lead to variations in the mean molecular weight of the atmosphere.
Recall that $R = R^*/M$, where $M$ is the mean molecular weight.  Since the simulations were done with dry air (molecular weight 29)
as the noncondensible and water vapor (molecular weight 18) as the condensible, the lower specific humidity towards the poles in cases
P30 and P5 leads to higher mean molecular weight and lower $R$ in polar regions, yielding an effect analogous to decreasing temperature
poleward with fixed $R$.  The effect could be captured in the conventional form of the thermal wind relation by defining a virtual
temperature $T_v$ such that $RT = R_d T_v$, where $R_d$ is the gas constant for the noncondensible background atmosphere.  Note that
the sign of the compositional thermal wind would be reversed if the condensible had higher molecular weight than the background,
e.g. for condensible $\mathrm{H_2O}$ or $\mathrm{CO_2}$ in noncondensible $\mathrm{H_2}$. In such situations, the moderately nondilute
climates would exhibit weaker or even easterly vertical shear. 

The intermediate diluteness cases exhibit a Hadley cell of the classic form, but it gets wider, deeper (as measured by range of
pressure it covers) and weaker as the nondiluteness is increased. The depth increase results from the elevation of the tropopause, which
in turn is associated with the weak lapse rate associated with hot moist adiabats.  The causes of the broadening are more subtle and require further analysis, which we will not pursue here. 

Unlike the pure steam case, the intermediate diluteness cases do exhibit transient eddies.  In case P5, the transience consists of
fairly large scale convective clusters that  appear and disappear without any clear propagation.  Though there is some eddy activity in
the midlatitudes, this is largely barotropic and almost all precipitation in case P5 is convective precipitation associated with the
tropical convective clusters. It is interesting that the compositional meridional baroclinicity does not give rise to baroclinic
instabilities. This may arise because ascending, condensing motions do not alter the mass distribution, while noncondensing descending motions expend
energy in subsiding through a statically stable atmosphere. The stability of jets arising from compositional pressure gradients requires
further study, and this, as well as an analysis of the tropical transient eddies, will be reserved for future work. 

\begin{figure}[!h]
\centering
\includegraphics[width=\textwidth]{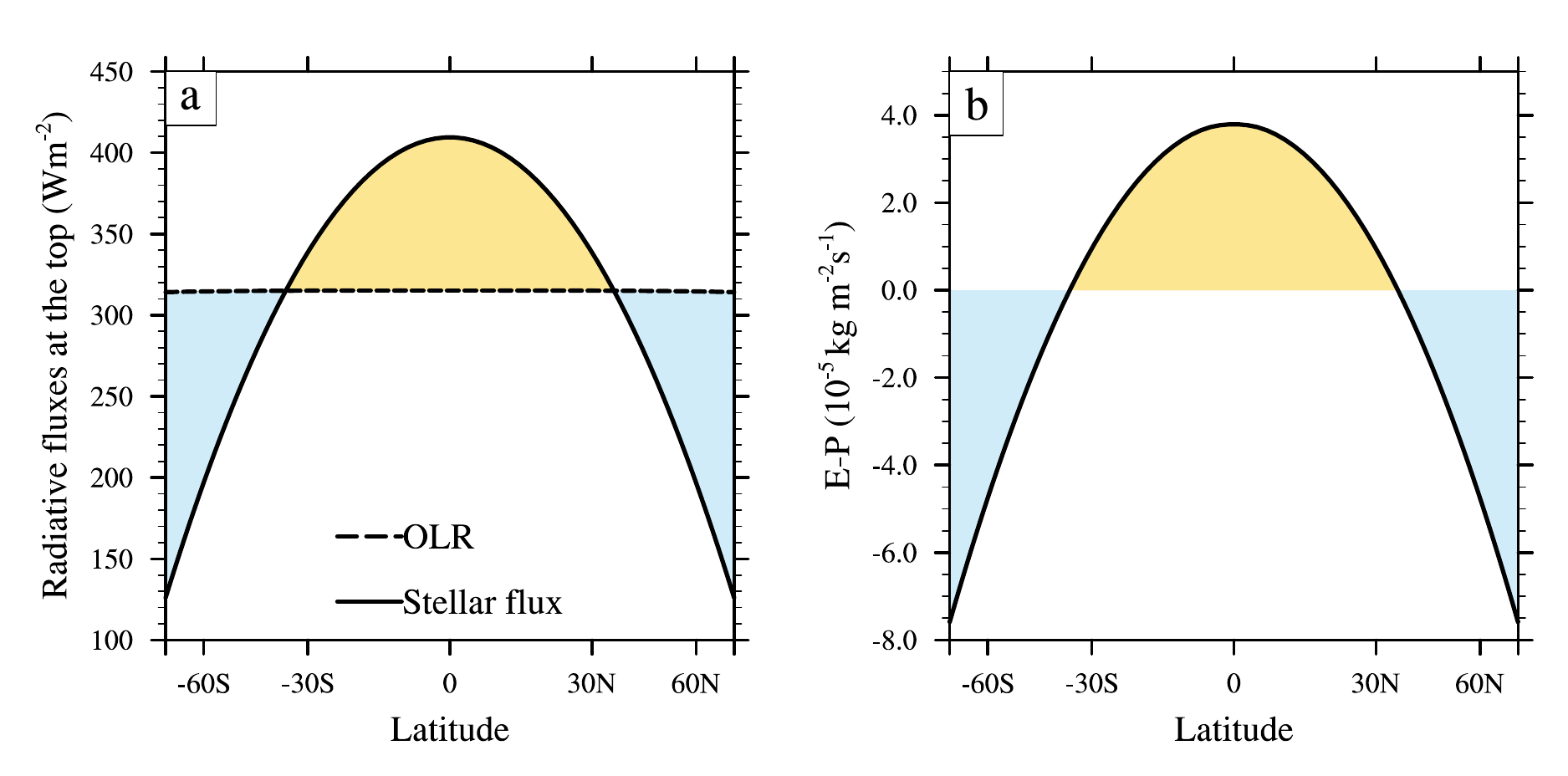}
\caption{(a) Meridional profiles of the insolation (solid) and the OLR (dashed) in experiment P0. The yellow shaded area marks the latitudes with radiative energy surplus, while the blue ones mark the latitudes with energy deficit. (b) Meridional profile of the net evaporation rate at the surface (E-P) in experiment P0. Similarly, the yellow shaded area shows the latitudes where net water mass flux goes from the surface reservoir to the atmosphere. The model has reached both mass and energy equilibrium, indicating that the area of the yellow shaded region is equal to the sum of the two blue ones in both two plots.}
\label{fig4_olr}
\end{figure}

Since the pole-to-equator instellation gradient is held fixed in these simulations, the weakening of the meridional temperature gradient
as nondiluteness is increased implies that the atmosphere must be carrying heat poleward at a greater rate in the more nondilute
cases. This may seem surprising, given that the atmospheric mass goes down in these simulations as the noncondensible background
gas is reduced.  However, while total atmospheric mass goes down, the mass of water vapor in the atmosphere rises markedly from
experiment P1000 to P0, due to the decrease in vertical lapse rate and the rise of atmospheric relative humidity.  This
allows increased poleward latent heat transport ($\int_0^{p_s} (L q v) \upd p/g$). leading to the uniform surface temperature in experiment P0. Figure~\ref{fig4_olr} shows the meridional distributions of the shortwave and longwave radiative fluxes at the top of the model together with $(E-P)$ for the pure steam case. The outgoing longwave radiation is essentially independent of latitude in this case.   $(E-P)$ has the right
shape and magnitude (when scaled by latent heat) to balance the radiative imbalance, verifying that, though thin, the latent heat of the
atmosphere allows it to nonetheless transport energy efficiently.  This mechanism breaks down for atmospheres where the temperature can only
produce an extremely low vapor pressure, as is the case for the local condensing $\mathrm{SO}_2$ atmosphere of Io. The analogous local rock vapor
atmospheres discussed in \cite{castan11} are not especially thin, nor is the latent heat of rock vapor condensation notably small; the
breakdown of the barotropic global atmosphere behavior in this case is probably due to the high rate of cooling of extremely hot atmospheres. 

\subsection{Saturation dynamics}
\label{section:SubsaturationDynamics}
 
The relative humidity of an air parcel is the ratio of the partial pressure of the condensible substance in that air parcel to the saturation
vapor pressure corresponding to the temperature of the air parcel. 
 
The relative humidity distribution affects the mass of water vapor in the atmosphere and therefore the meridional heat transport. Further, when the
condensible is a greenhouse gas (as in the case of water) it has a strong impact on the radiation budget of the planet. For Earth's atmosphere, how the relative humidity is distributed and how this distribution changes in a warming world are key questions involved in understanding the feedbacks that determine the sensitivity of climate to changes in radiative forcing. This would be equally true for other planets for which the condensible
substances have a significant greenhouse effect.

A paradigm of nonlocal large-scale control of tropospheric humidity has long been proposed for understanding the relative humidity distribution in Earth's atmosphere, known as the advection-condensation model \citep{pierrehumbert1998lateral,pierrehumbert07}. This model makes use of the 
property that the mixing ratio of the condensible in an air parcel is conserved along any segment of a trajectory which doesn't encounter 
conditions where it becomes saturated and loses water by precipitation.  Hence the relative humidity of the air parcel at its destination is the ratio of the saturation molar mixing ratio $r_{sat}$ of water vapor at the position of the last saturation to that at the destination, 
and the relative humidity at the final position has no sensitivity to the air parcel's history at times earlier than the last saturation.
This is an idealized model of relative humidity, as unsaturated air parcels can gain moisture by evaporation of precipitation or mixing with
neighboring moist parcels, or lose it by mixing with neighboring dry parcels; however the advection-condensation model has been found to
have considerable quantitative explanatory value.  In terms of equations, the last-saturation model giving the relative humidity
at position $\mathbf{x}$ can be expressed as
\begin{equation}
RH(\mathbf{x}) \equiv \frac{p_c(\mathbf{x})}{p_\mathrm{sat}(\mathbf{x})} 
                        =\frac{r_c(\mathbf{x})}{r_\mathrm{sat}(\mathbf{x})}
                        = \frac{r_\mathrm{sat}(\mathbf{x}_\mathrm{last})}{r_\mathrm{sat}(\mathbf{x})} 
                        = \frac{p_\mathrm{sat}(\mathbf{x}_{last})}{p_\mathrm{sat}(\mathbf{x})  } \frac{p_a(\mathbf{x}) }{p_a(\mathbf{x}_{last}) }
\label{eqn:LastSat}
\end{equation}
where $r_c \equiv p_c/p_a$ is the molar mixing ratio of the condensible, $r_\mathrm{sat}$ its value at saturation, $p_\mathrm{sat}$ is the saturation vapor pressure and the subscript `last'  stands for the position of last saturation. This model was used to reconstruct relative humidity in the Earth's subtropics using a backward Lagrangian trajectory technique, and showed excellent agreement with the Meteosat satellite observations \citep{pierrehumbert1998lateral}. The assumption about water mass in an air parcel in the advection-condensation model is exactly what occurs in our GCM since there is no microphysical cloud scheme or small-scale unresolved turbulent mixing (apart from numerical diffusion). Therefore we can use the concept of last saturation to explain the simulated relative humidity distribution. The typical situation leading to strong subsaturation is when
the air parcel came from a higher, colder place. In that case, the first ratio at the end of Eq. \ref{eqn:LastSat} is very small, because
of the exponential dependence of saturation vapor pressure on temperature; the second ratio is greater than unity because of the subsidence and
compression of the air parcel, but it is invariably overwhelmed by the exponentially small first term. 

\begin{figure}[!h]
\centering
\includegraphics[width=\textwidth]{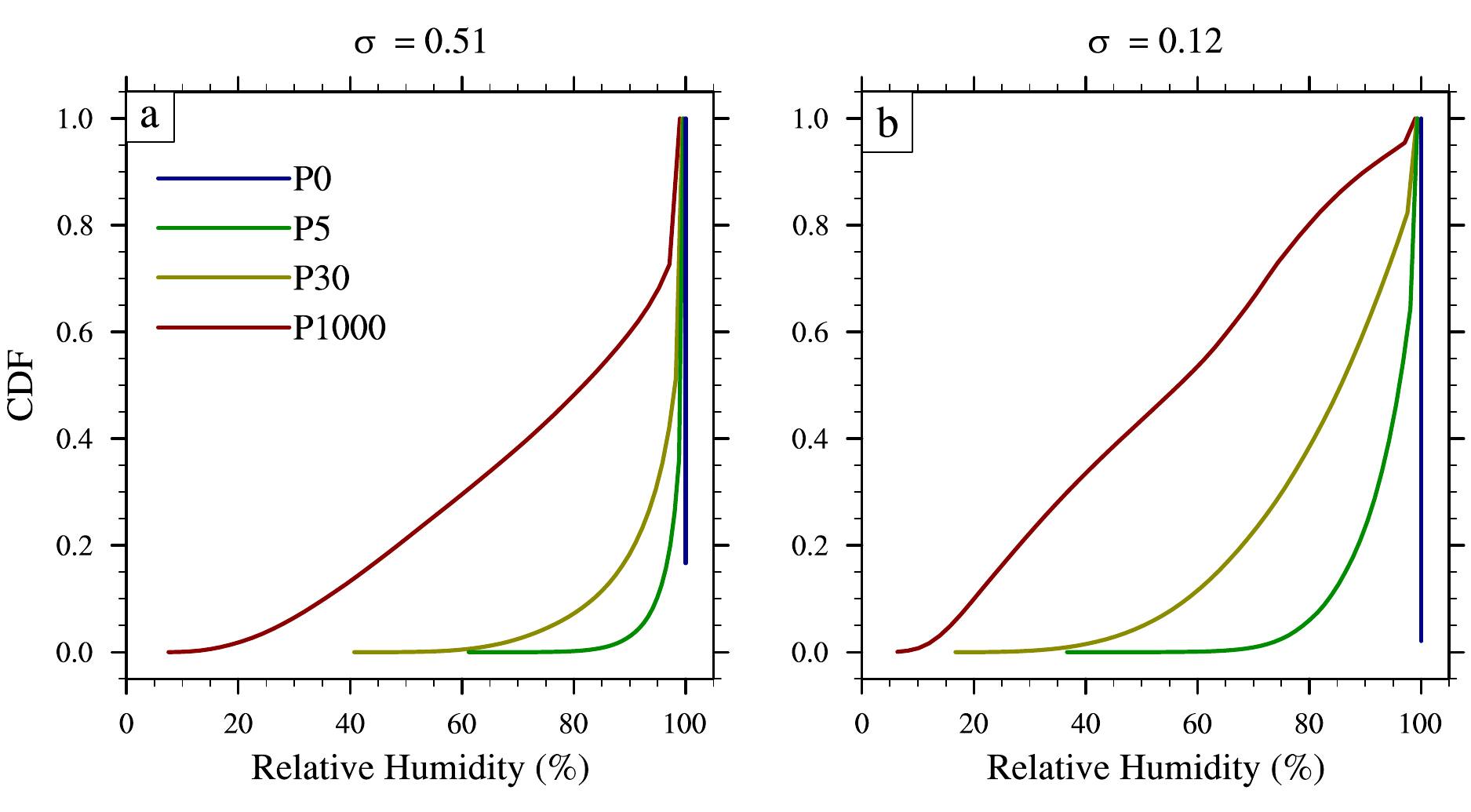}
\caption{Cumulative distribution function of relative humidity in experiments P1000 (red), P30 (yellow), P5 (green), P0 (blue) on the $\sigma$ = 0.51 (a), 0.12 (b) surfaces. The data samples are taken from a 100-day integration and collected once per day.}
\label{fig5_pdf}
\end{figure}

Figure~\ref{fig5_pdf} shows the cumulative distribution of the relative humidity for the four simulations. Results are shown on two different $\sigma$ surfaces: one near the mid-troposphere ($\sigma = 0.51$) and the other in the upper atmosphere ($\sigma = 0.12$).  First, because the saturation mixing ratio decreases with height, the driest air parcel in the troposphere usually forms in the subtropics and is last saturated at the coldest location in the troposphere--the tropopause at the equator. We can roughly estimate the minimum relative humidity in experiment P1000 by the advection-condensation model. From the mass steamfunction in Figure~\ref{fig3_upsi}e and the air temperature in Figure~\ref{fig2_tq}a, the driest air parcel should be last saturated near 150\,hPa with the temperature of 235.4\,K. Since the subtropics also satisfy the WTG approximation, the temperature at 510\,hPa can be computed by the moist adiabat, which is 283.4\,K. Then the estimated relative humidity given by Eq.(\ref{eqn:LastSat}) is 7.0\%, and very close to the minimum value in the simulation (the left end point of the red curve in Figure~\ref{fig5_pdf}a). Second, Figure~\ref{fig5_pdf}a and b show  a general trend that the troposphere becomes more saturated as the water vapor concentration rises. This is again related to the decrease in the vertical lapse rate $(\upd \ln T/\upd \ln p)$, which leads to a decrease in the vertical relative change of saturation mixing ratio $(\upd r_{sat}/\upd p)$ and a small expansion of the troposphere (Figure~\ref{fig2_tq}a-c). Although the latter one tends to dry the upper troposphere, the former one wins the competition. 

The role of the last-saturation statistic changes markedly, however, as the pure-steam limit is approached.  For a nearly pure-steam 
atmosphere, $p_\mathrm{sat}(\mathbf{x}_{last}) \approx p(\mathbf{x}_{last})$ and since $p_a/p_c$ is conserved along the noncondensing
trajectory $p_a(\mathbf{x}_{last})/p_c(\mathbf{x}_{last}) = p_a(\mathbf{x})/p_c(\mathbf{x})$, which in the nearly pure steam limit
implies $p_a(\mathbf{x})/p_a(\mathbf{x}_{last}) \approx p(\mathbf{x})/p(\mathbf{x}_{last})$. Plugging this into
the last equality in Eq. \ref{eqn:LastSat} then implies
\begin{equation}
RH(\mathbf{x}) = \frac{p(\mathbf{x})}{p_\mathrm{sat}(\mathbf{x})}
\end{equation}
This depends only on the local temperature and pressure where the parcel lands. It is independent of the parcel's history, except
insofar as that history affects the final temperature of the parcel. In contrast to the more dilute cases, subsaturation cannot
be created by "bringing down" dry air from aloft. It can only create subsaturation by heating up an air parcel, either by adiabatic
compression or net radiative heating. In our simulation there is no net radiative heating internal to the atmosphere, but in principle,
subsaturated air could still be produced by subsiding motions that are fast enough to overcome radiative cooling. This does not happen
to any significant extent in our pure steam simulation (the two blue lines in Figure~\ref{fig5_pdf}a and b), and the atmosphere to all intents and purposes becomes saturated
everywhere except in the very thin stratosphere (barely visible in the figure) where radiative
heating warms the atmosphere to temperatures greater than the dew point. 

\begin{figure}[!h]
\centering
\includegraphics[width=\textwidth]{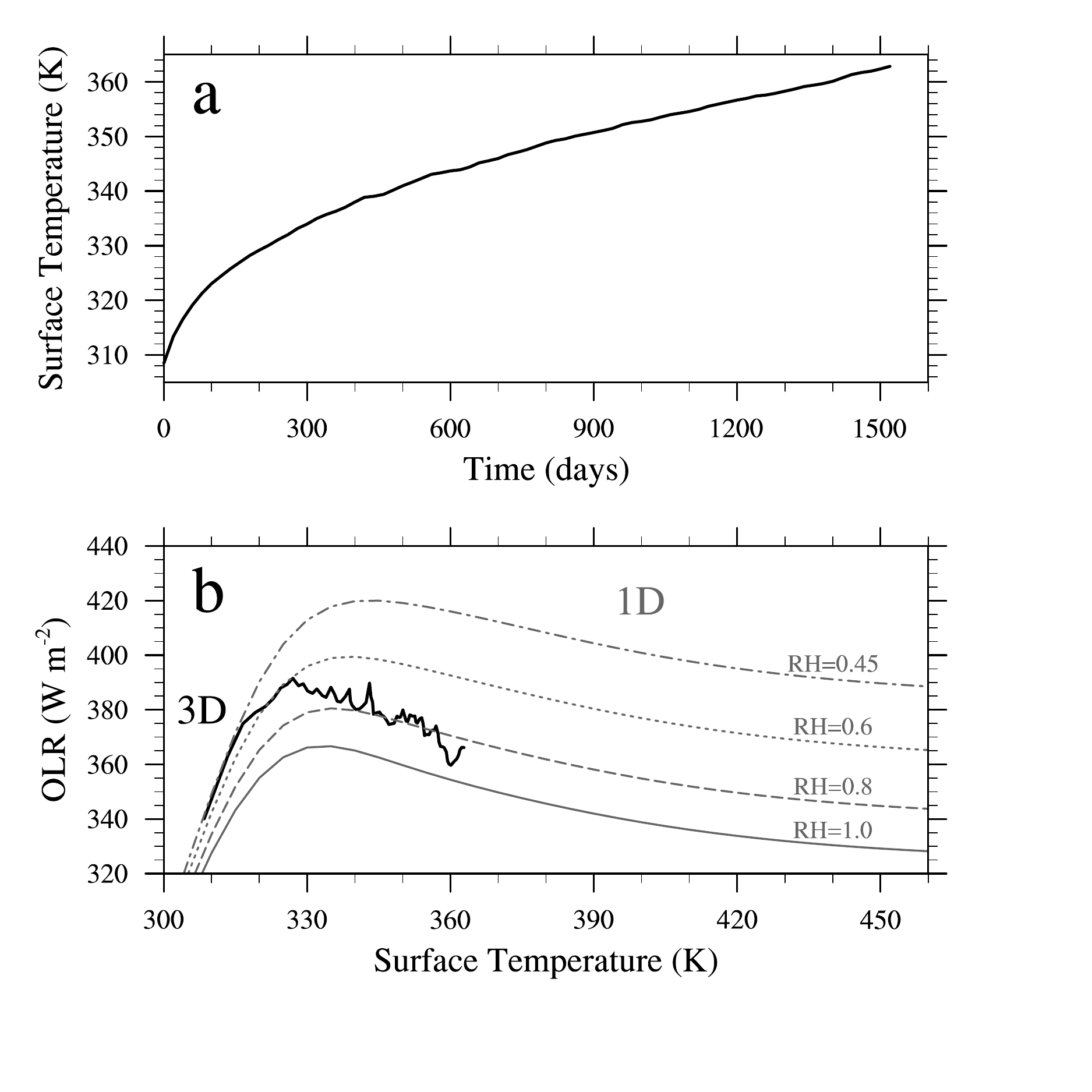}
\caption{(a) Time evolution of the tropical mean $T_s$ (averages are taken within 30$^\circ$ of latitude) in the runaway greenhouse experiment with global mean insolation of 450\,W\,m$^{-2}$. (b) OLR dependence on $T_s$ in tropics. The black curve shows results for the idealized 3D simulation, and gray curves show results for for 1D radiative-convective simulations. The 1D model assumes moist adiabat and uses various values of relative humidity in the 1D radiative transfer. }
\label{fig6_runaway}
\end{figure}

The upper tropospheric relative humidity (UTRH) can vary the pressure of the radiating level (the optical thickness between the radiating level and the top of the atmosphere is approximately unity for an optically thick gray atmosphere), and therefore the OLR.  
It is not easy to show the change of the relative humidity above the radiating level directly in the simulations listed in Table~\ref{table_pdry}. 
Here we develop another experiment and simulate a runaway greenhouse atmosphere to study  the change of the UTRH as the atmosphere becomes non-dilute and the radiative effect due to this change. We use the temperature, humidity and wind fields on day 3000 of the experiment P1000 as the initial condition and trigger a runaway greenhouse by raising the global mean insolation to 450\,W\,m$^{-2}$. The value is well above the threshold of 1D runaway greenhouse assuming a relative humidity of 45\% in our gray radiation scheme. We plot the OLR as a function of $T_s$ with both quantities averaged in the tropics in Figure~\ref{fig6_runaway}b. Due to the WTG approximation, the comparison between the modeling result and the 1D reference curves corresponding to various relative humidities could roughly\footnote{The relation between the OLR and the water path is not linear} give the degree of saturation averaged near the tropical  radiating level. The relative humidity first stays constant at 45\% until the $T_s$ reaches 315\,K, and then it continue to rise towards unity as the planet warms. Another idealized simulation on the 3D runaway greenhouse gray atmosphere  also show similar trend of the UTRH (Figure~7 in \cite{ishiwatari02}).  Recently,   \citet{leconte13}  conducted a study of the runaway greenhouse in an idealized 3D GCM with a real-gas radiation scheme. This new simulation showed a decrease in UTRH before runaway greenhouse occurs ($T_s \approx 320\,$K, Extended Data Figure~3b in \cite{leconte13}), probably because of the emergence of a cold tropopause owing to real-gas radiation effects.  However, when the surface temperature continues to increase and the runaway greenhouse atmosphere becomes non-dilute, the UTRH should rise again and approach unity according to our discussion regarding $\upd r_\mathrm{sat} / \upd p$ above. This effect is not seen in \cite{leconte13}, but some indication of this behavior is found in the simulation of the same problem in \cite{wolf2015evolution},
which also edges into the nondilute regime.

This increase in UTRH leads to an interesting question about multiple equilibria in the real runaway greenhouse atmosphere. Assuming constant planetary albedo that does not change with $T_s$, for a pure water-vapor atmosphere, there is only one stable equilibrium state when the net absorbed stellar flux $(S_{abs})$ is less than the Komabayashi-Ingersol limit of the water vapor atmosphere ($F_{KI} \approx 282\,$W\,m$^{-2}$), and no equilibrium state when $S_{abs} > F_{KI}$. For the Earth's atmosphere, the 1000\,hPa of non-condensible components not only dilutes the atmosphere, but also leads to the sub-saturation of the upper troposphere and therefore increases the maximum amount of energy the planet could emit into space ($F_{max}\sim 320\,$W\,m$^{-2}$ shown in Extended Data Figure~3b in \cite{leconte13}). Hence, two equilibrium states exists when $F_{KI} < S_{abs} < F_{max} $: one is stable and another unstable. 
In a real water-vapor atmosphere, the planetary albedo depends on $T_s$ as well. Le Conte \textit{et al.}  show a downward trend of the planetary albedo in a dilute atmosphere as the planet warms\citep{leconte13}. For a non-dilute atmosphere, the planetary albedo is primarily determined by the Mie scattering of water clouds and Rayleigh scattering and absorption of water molecules. The latter effect can be easily computed, but the former one is not for we still know little about the cloud microphysics when the atmosphere is non-dilute. If the planetary albedo is a strongly non-linear function of $T_s$, then more equilibria (possibly stable) may exist in the non-dilute atmosphere. In fact, when applying the relation between the planetary albedo and $T_s$ estimated in a 1D model (Figure~3b in \cite{kopparapu13}), there would be another stable equilibrium state with $ T_s \sim 400\,$K under our current insolation.   \citet{goldblatt15} carried out a similar analysis and showed both a cold ($270 \lessapprox T_s \lessapprox 290\,$K) and hot ($350 \lessapprox T_s \lessapprox 550\,$K) stable climate state for a pure water atmosphere.

\section{Discussion}
\label{section:discussion}
 
Given the vast amount of energy stored as latent heat in nondilute atmospheres, one might expect 
them to exhibit violent behavior, with extreme convective storms perhaps reducing pressures
locally to a near vacuum as massive proportions of the atmosphere rain out.  In fact, we
have found strongly nondilute atmospheres to be rather quiescent places, with largely
barotropic dynamics, weak temperature gradients, sluggish circulations, and rather gentle
laminar ascent in convective regions instead of the boiling cumulus towers familar
from the Earth's tropics.  All of these features can be understood in terms of
well-established thermodynamic and GFD principles applied in a novel way. 
Ironically, it may be difficult or impossible for hurricanes to form in highly nondilute atmospheres.
The quiescent nature of nondilute atmospheres is already suggested by the example of Titan, which
has rather weak condensationally-driven activity in its troposphere despite the very
considerable amount of energy stored in the form of latent heat of condensible methane --
vast amounts of energy are there, but the dynamics can't get at it. Titan, however,
is weakly illuminated and slowly rotating, which may also play a role in its
quiescent (though nonetheless interesting) dynamics. We have shown that similar
behavior emerges even in rapidly rotating cases with Earthlike instellation. 

An important result obtained from our simulations is that atmospheres become
increasingly saturated throughout a deepening layer as they become more
nondilute.  This has critical implications for the conditions which support
a runaway greenhouse for water vapor or other condensible greenhouse gases,
as dilute atmospheres can be inhibited from entering into a runaway by 
subsaturation.  The increasing saturation with temperature we have found
also confirms earlier speculations that the feedback of increasing
relative humidity can potentially lead to multiple equilibria, in which
atmospheres can have a metastable non-runaway state and a hot runaway state
for the same instellation conditions (though our simulations do not themselves
exhibit this multiplicity). An important implication of our results on saturation
is that a substantial inventory of noncondensing background gas helps to protect
a planet against a runaway greenhouse, through allowing significant subsaturation.
This effect is of interest not just for water oceans, but also for the
preservation of $\mathrm{CO_2}$ oceans, which are of current interest as 
a possible alternative life-supporting environment \citep{KISSNotJustWater}.

The tendency toward saturation should also make clouds more prevalent in 
nondilute atmospheres, as clouds form in saturated or supersaturated air -- though
the weakness of vertical motions in nondilute atmospheres could work against
this expectation. 


Though in some sense sluggish, nondilute atmospheres nonetheless exhibit a variety
of novel and challenging dynamical phenomena.  Angular momentum carried by
precipitation allows global or near-global surface westerlies to prevail.
One can raise related questions about the way nondilute precipitation 
affects potential vorticity dynamics.  
Compositionally induced pressure gradients allow strong baroclinic jets
even when horizontal temperature gradients are weak, but these jets
appear to be stable against baroclinic instability. The width of the
weak nondilute Hadley circulations remains to be understood, and
the tropics of nondilute atmospheres exhibit a variety of large scale
transient convective clusters whose dynamics remain to be elucidated;
they are likely to have a strong influence on the cloud patterns of
nondilute atmospheres. 

This paper also serves to introduce a new tool -- Exo-FMS -- we have developed for simulation of the 3D climate dynamics of 
quite general exoplanet atmospheres.  This model will be put to further use in future work, and it is our intention to develop it
into an open-source platform for planetary circulation modeling.  Future studies will include the examination of nondilute
circulations on tide-locked planets, study of planets where most of the instellation is deposited in the atmosphere rather than at
a distinct surface, and characterization of a range of possible cloud effects. More generally, there is considerable
work that needs to be done in order to incorporate microphysical effects into the modeling of nondilute atmospheres.
In the absence of suitable cloud condensation nuclei, atmospheres may become highly supersaturated before condensate
forms, and this is bound to have dynamical effects.  Further, the effect of retained condensate in the case where 
condensate particles remain small can significantly affect buoyancy and inhibit convection. The microphysics governing
precipitation and re-evaporation also affect the vertical transport of chemical species in a planetary atmosphere, and
therefore affect the vertical structure of atmospheric composition. 

Few of the dynamical aspects we have discussed are currently accessible to detailed observational study.
Observational techniques, however, continue to advance at a breathtaking pace. This will expand the range
of theories that can be tested. Inevitably, such advances will also turn up additional surprises. 
We should be prepared to think about them, and that will require climate dynamics thinking well outside
the terrestrial box. Nondilute atmospheres are but one example of the class of novel atmospheres that
will need to be considered. 

\section*{Acknowledgment}

We thank Robin Wordsworth, Dorian Abbot, Daniel Koll and Jun Yang for many helpful discussions. Support for this work was provided by the NASA Astrobiology Institute Virtual Planetary Laboratory Lead Team, under the National Aeronautics and Space Administration solicitation NNH12ZDA002C and Cooperative Agreement Number NNA13AA93A.

\bibliography{NonDilute}

\begin{thebibliography}{}
\expandafter\ifx\csname natexlab\endcsname\relax\def\natexlab#1{#1}\fi

\bibitem[{Ackerman \& Marley(2001)}]{ackerman2001precipitating}
Ackerman, A.~S., \& Marley, M.~S. 2001, The Astrophysical Journal, 556, 872

\bibitem[{Barshay \& Lewis(1976)}]{BarshayLewis1976}
Barshay, S.~S., \& Lewis, J.~S. 1976, ARA\&A, 14, 81

\bibitem[{Batalha(2014)}]{batalha2014exploring}
Batalha, N.~M. 2014, Proceedings of the National Academy of Sciences, 111,
  12647

\bibitem[{Castan \& Menou(2011)}]{castan11}
Castan, T., \& Menou, K. 2011, ApJ, 743, L36

\bibitem[{Cooper {et~al.}(2003)Cooper, Sudarsky, Milsom, Lunine, \&
  Burrows}]{cooper2003modeling}
Cooper, C.~S., Sudarsky, D., Milsom, J.~A., Lunine, J.~I., \& Burrows, A. 2003,
  The Astrophysical Journal, 586, 1320

\bibitem[{Ding \& Pierrehumbert(2016)}]{FengNondiluteColumnModel}
Ding, F., \& Pierrehumbert, R. 2016, The Astrophysical Journal, 822, 24

\bibitem[{Emanuel(1994)}]{emanuel}
Emanuel, K.~A. 1994, Atmospheric {Convection} (Oxford University Press)

\bibitem[{Frierson(2007)}]{frierson}
Frierson, D. M.~W. 2007, J. Atmos. Sci., 64, 1959

\bibitem[{Goldblatt(2015)}]{goldblatt15}
Goldblatt, C. 2015, Astrobiology, 15, 362

\bibitem[{Goldblatt {et~al.}(2009)Goldblatt, Claire, Lenton, Matthews, Watson,
  \& Zahnle}]{goldblatt09}
Goldblatt, C., Claire, M.~W., Lenton, T.~M., {et~al.} 2009, Nature Geosci, 2,
  891

\bibitem[{Heng \& Showman(2015)}]{HengShowmanAREPS2015}
Heng, K., \& Showman, A.~P. 2015, Annual Review of Earth and Planetary
  Sciences, 43, 509

\bibitem[{Ingersoll(1990)}]{ingersoll1990Triton}
Ingersoll, A.~P. 1990, Nature, 344, 315

\bibitem[{Ingersoll {et~al.}(1985)Ingersoll, Summers, \& Schlipf}]{ingersoll85}
Ingersoll, A.~P., Summers, M.~E., \& Schlipf, S.~G. 1985, Icarus, 64, 375

\bibitem[{Ishiwatari {et~al.}(2002)Ishiwatari, Takehiro, Nakajima, \&
  Hayashi}]{ishiwatari02}
Ishiwatari, M., Takehiro, S.-i., Nakajima, K., \& Hayashi, Y.-Y. 2002, Journal
  of the Atmospheric Sciences, 59, 3223

\bibitem[{Joshi(2003)}]{joshi2003TideLock}
Joshi, M. 2003, Astrobiology, 3, 415

\bibitem[{Koll \& Abbot(2016)}]{KollAbbotApJ2016}
Koll, D. D.~B., \& Abbot, D.~S. 2016, Astrophysical Journal, , in press

\bibitem[{Kopparapu {et~al.}(2013)Kopparapu, Ramirez, Kasting, Eymet, Robinson,
  Mahadevan, Terrien, Domagal-Goldman, Meadows, \& Deshpande}]{kopparapu13}
Kopparapu, R.~K., Ramirez, R., Kasting, J.~F., {et~al.} 2013, ApJ, 765, 131

\bibitem[{Leconte {et~al.}(2013)Leconte, Forget, Charnay, Wordsworth, \&
  Pottier}]{leconte13}
Leconte, J., Forget, F., Charnay, B., Wordsworth, R., \& Pottier, A. 2013,
  Nature, 504, 268

\bibitem[{Lin(2004)}]{lin}
Lin, S.-J. 2004, Mon. Wea. Rev., 132, 2293

\bibitem[{Lorenz {et~al.}(1999)Lorenz, McKay, \&
  Lunine}]{lorenz1999TitanStability}
Lorenz, R.~D., McKay, C.~P., \& Lunine, J.~I. 1999, Planetary and space
  science, 47, 1503

\bibitem[{Madhusudhan {et~al.}(2012)Madhusudhan, Lee, \&
  Mousis}]{madhusudhanCarbon2012}
Madhusudhan, N., Lee, K.~K., \& Mousis, O. 2012, The Astrophysical Journal
  Letters, 759, L40

\bibitem[{Manabe \& Strickler(1964)}]{manabe}
Manabe, S., \& Strickler, R.~F. 1964, Journal of the Atmospheric Sciences, 21,
  361

\bibitem[{Markiewicz {et~al.}(2007)Markiewicz, Titov, Limaye, Keller, Ignatiev,
  Jaumann, Thomas, Michalik, Moissl, \& Russo}]{markiewicz2007VEXWinds}
Markiewicz, W., Titov, D., Limaye, S., {et~al.} 2007, Nature, 450, 633

\bibitem[{Merlis \& Schneider(2010)}]{merlis}
Merlis, T.~M., \& Schneider, T. 2010, J. Adv. Model. Earth Syst., 2, 13

\bibitem[{O'Gorman \& Schneider(2008)}]{ogorman}
O'Gorman, P.~A., \& Schneider, T. 2008, J. Climate, 21, 3815

\bibitem[{Pierrehumbert(1998)}]{pierrehumbert1998lateral}
Pierrehumbert, R. 1998, Geophysical research letters, 25, 151

\bibitem[{Pierrehumbert {et~al.}(2011)Pierrehumbert, Abbot, Voigt, \&
  Koll}]{pierrehumbert2011AREPSNeoprot}
Pierrehumbert, R., Abbot, D., Voigt, A., \& Koll, D. 2011, Annual Review of
  Earth and Planetary Sciences, 39, 417

\bibitem[{Pierrehumbert {et~al.}(2007)Pierrehumbert, Brogniez, \&
  Roca}]{pierrehumbert07}
Pierrehumbert, R., Brogniez, H., \& Roca, R. 2007, in The Global Circulation of
  the Atmosphere, ed. T.~Schneider \& A.~H. Sobel (Princeton University Press),
  143--185

\bibitem[{Pierrehumbert \& Gaidos(2011)}]{pierrehumbert2011hydrogen}
Pierrehumbert, R., \& Gaidos, E. 2011, The Astrophysical Journal Letters, 734,
  L13

\bibitem[{Pierrehumbert(2011{\natexlab{a}})}]{pierrehumbert2011palette}
Pierrehumbert, R.~T. 2011{\natexlab{a}}, The Astrophysical Journal Letters,
  726, L8

\bibitem[{Pierrehumbert(2011{\natexlab{b}})}]{ClimateBook}
---. 2011{\natexlab{b}}, {Principles of Planetary Climate} (Cambridge
  University Press)

\bibitem[{Pierrehumbert(2013)}]{pierrehumbert2013strange}
---. 2013, Nature Geoscience, 6, 81

\bibitem[{Pierrehumbert \& Lloyd(2016)}]{KISSNotJustWater}
Pierrehumbert, R.~T., \& Lloyd, J. 2016, in Don't Follow (Just) the Water: Does
  Life Occur in Non-Aqueous Media {, Keck Inst. for Space Studies} Report (in
  prep.), ed. J.~B. Chris~Sotin \& J.~Lunine (Keck Institute for Space Studies)

\bibitem[{Showman {et~al.}(2013)Showman, Wordsworth, Merlis, Kaspi, Bullock, \&
  Harder}]{showman2013atmospheric}
Showman, A.~P., Wordsworth, R.~D., Merlis, T.~M., {et~al.} 2013, Comparative
  Climatology of Terrestrial Planets, 1, 277

\bibitem[{Sing {et~al.}(2015)Sing, Fortney, Nikolov, Wakeford, Kataria, Evans,
  Aigrain, Ballester, Burrows, Deming,
  {et~al.}}]{SingClearToCloudyHotJupitersNature2015}
Sing, D.~K., Fortney, J.~J., Nikolov, N., {et~al.} 2015, Nature

\bibitem[{Som {et~al.}(2012)Som, Catling, Harnmeijer, Polivka, \&
  Buick}]{som12}
Som, S.~M., Catling, D.~C., Harnmeijer, J.~P., Polivka, P.~M., \& Buick, R.
  2012, Nature, 484, 359

\bibitem[{Vallis(2006)}]{vallis}
Vallis, G.~K. 2006, Atmospheric and {Oceanic} {Fluid} {Dynamics}:
  {Fundamentals} and {Large}-scale {Circulation} (Cambridge University Press)

\bibitem[{Williams {et~al.}(2009)Williams, Pierrehumbert, \&
  Huber}]{FalseThermostats2009}
Williams, I.~N., Pierrehumbert, R.~T., \& Huber, M. 2009, Geophysical Research
  Letters, 36

\bibitem[{Wolf \& Toon(2015)}]{wolf2015evolution}
Wolf, E., \& Toon, O. 2015, Journal of Geophysical Research: Atmospheres, 120,
  5775

\bibitem[{Wordsworth \& Pierrehumbert(2013)}]{wordsworth13}
Wordsworth, R., \& Pierrehumbert, R. 2013, Science, 339, 64

\bibitem[{Wordsworth \& Pierrehumbert(2014)}]{wordsworth2014abiotic}
---. 2014, The Astrophysical Journal Letters, 785, L20

\bibitem[{Wordsworth {et~al.}(2011)Wordsworth, Forget, Selsis, Millour,
  Charnay, \& Madeleine}]{wordsworth2011gliese}
Wordsworth, R.~D., Forget, F., Selsis, F., {et~al.} 2011, The Astrophysical
  Journal Letters, 733, L48

\end{thebibliography}
\bibliographystyle{apj}
%
%
%
%
%

\end{document}